\documentclass[12pt]{article}

\usepackage{graphics,epsfig, psfrag}
\def \HT{{\mathcal H}}
\def \LT{{\mathcal L}}
\def \HCT{{\hat{\mathcal H}}}

\def \s{\scriptscriptstyle}

\begin{document}

\makeatletter
\@addtoreset{equation}{section}
\makeatother
\renewcommand{\theequation}{\thesection.\arabic{equation}}

\begin{titlepage}

\baselineskip =15.5pt
\pagestyle{plain}
\setcounter{page}{0}

\begin{flushright}
%work in progress
\end{flushright}
\vfil

\begin{center}
{\huge Minimal Coupling in Koopman-von Neumann Theory}
\end{center}

\vfil

\begin{center}
{\large E. Gozzi}\footnote{e-mail: gozzi@ts.infn.it} and {\large D. Mauro}\footnote{e-mail: mauro@ts.infn.it}\\
\vspace {1mm}
Dipartimento di Fisica Teorica, Universit\`a di Trieste, \\
Strada Costiera 11, P.O.Box 586, Trieste, Italy \\ and INFN, Sezione 
di Trieste.\\
\vspace {1mm}
\vspace{3mm}
\end{center}

\vfil

\noindent
Classical mechanics (CM), like quantum mechanics (QM), can have an {\it operatorial} formulation.
This was pioneered by Koopman and von Neumann (KvN) in the 30's. They basically
formalized, via the introduction of a classical {\it Hilbert space}, earlier work of Liouville 
who had shown that the classical time evolution
can take place via an operator, nowadays known as the Liouville operator.
In this paper we study how to perform the coupling of a point particle to a gauge field in the 
KvN version of CM. So we basically implement at the classical operatorial level
the  analog of the minimal coupling of QM. We show 
that, differently than in  QM, not only the momenta but also other variables have to be coupled to 
the gauge field. We also analyze in details how the gauge invariance manifests itself in the Hilbert space 
of KvN and indicate the differences with QM. As an application of the KvN method we study  
the Landau problem proving that
there are  many more degeneracies at the classical operatorial level than at the quantum one. As a second example we 
go through the  Aharonov-Bohm phenomenon showing that, at the quantum level, this phenomenon
manifests its effects on the spectrum of the quantum Hamiltonian while at the classical level there is no
effect whatsoever on the spectrum of the  Liouville operator.

\vfil
\end{titlepage}
\newpage

\section{Introduction}

It is well-known that in {\it classical} statistical mechanics the evolution of probability densities $\rho(q,p)$ is given
by the Liouville equation
%%%
\begin{equation}
\partial_t\rho(q,p)=-\widehat{L}\rho(q,p) \label{lio}
\end{equation}
%%%
where $\widehat{L}=\partial_pH\partial_q-\partial_qH\partial_p$  and $H(q,p)$ is the Hamiltonian of the system. The 
$\rho(q,p)$ must only be integrable functions (i.e. $\rho\in L^1$-space) because of their meaning as probability densities 
$(i.e. \int\rho\;dqdp<\infty$). As they are only $L^1$-functions they do not make up an Hilbert space. In 1931 
KvN~\cite{Koopman} 
{\it postulated} 
the same evolution equation for complex distributions $\psi(q,p)$ making up an $L^2$ Hilbert space:
%%%
\begin{equation}
\partial_t\psi(q,p)=-\widehat{L}\psi(q,p) \label{lio2}
\end{equation}
%%%
If we postulate eq. (\ref{lio2}) for 
$\psi(q,p)$, then it is easy to prove that functions $\rho$ of the form:
%%%
\begin{equation}
\rho=|\psi|^2 \label{pro}
\end{equation}
%%%
evolves with the same equation as $\psi$. This is so because the operator $\widehat{L}$ contains only first order 
derivatives. This is not what happens in QM where the evolution of the $\psi(q)$ is via the Schr\"odinger operator
$\widehat{H}$ while the one of the associated $\rho=|\psi|^{2}$ is via a totally different operator. The reason
is that the Schr\"odinger operator $\widehat{H}$, differently than the Liouville operator
$\widehat{L}$, contains second order derivatives. By postulating the relations 
(\ref{pro}) and (\ref{lio2}) for the $\psi$, 
KvN managed to build an operatorial formulation for classical mechanics (CM) equipped with an Hilbert space structure and 
producing the same results as the Liouville formulation. Of course there are phases in a complex $\psi$ and one could
wonder which is their role in CM. This problem is addressed in ref.\cite{Mauro}.
 We will briefly review the KvN formalism in {\bf section 2}
of this paper. 
\par
The question we want to address in this paper is how to couple a gauge field to the point particle degrees
of freedom once we work in the operatorial formulation of CM. We know that in QM there is a simple rule known as {\it "minimal 
coupling"} (MC) \cite{quantumbooks}:
%%%
\begin{equation}
\displaystyle
\vec{p}\;\longrightarrow\;\vec{p}-\frac{e}{c}\vec{A} \label{mincoup}
\end{equation}
%%%
where $\vec{A}$ is the gauge field. This  rule says that it is enough to replace $\vec{p}$ with $\displaystyle
\vec{p}-\frac{e}{c}\vec{A}$ in the Hamiltonian
$\displaystyle H(q,p)\longrightarrow H(q,p-\frac{e}{c}A)$ in order to get the interaction of the particle with the 
gauge field and then represent $\vec{p}$ operatorialy as $\displaystyle \frac{\hbar}{i}\frac{\partial}{\partial x}$.
Since the operatorial formulation of CM has an Hilbert space structure like QM and it has operators like
$\widehat{L}$ analog of the Schr\"odinger $\widehat{H}$ operator of QM, we would like to find out the
"minimal coupling" rules which would transform the $\widehat{L}$ without a gauge field interaction into the one
$\widehat{L}_{\s A}$ with interaction.

Let us start by just using the rule (\ref{mincoup}) inside the $H$ which appears in $\widehat{L}$:
%%%
\begin{equation}
\widehat{L}=(\partial_pH)\partial_q-(\partial_qH)\partial_p \label{liuvillian}
\end{equation}
%%%
without modifying the derivative $\displaystyle \frac{\partial}{\partial p}$. We can work out the simple case of a particle moving
under a constant magnetic field directed along $z$. Let us choose the gauge field as:
%%%
\begin{equation}
\left\{
	\begin{array}{l}
	\displaystyle A_x=0\\
	\displaystyle A_y=Bx\\
	\displaystyle A_z=0\\
	\end{array}
	\right.
\end{equation}
%%%
Using the MC (\ref{mincoup}):
%%%
\begin{equation}
\displaystyle
p_y\;\longrightarrow\;p_y-\frac{eB}{c}x \label{pipsilon}
\end{equation}
%%%
the Hamiltonian  becomes
%%%
\begin{equation}
\displaystyle
H_{\s A}=\frac{p_x^2}{2m}+\frac{1}{2m}\biggl(p_y-\frac{eB}{c}x\biggr)^2+\frac{p_z^2}{2m}
\end{equation}
%%%
and $\widehat{L}$ would turn into
%%%
\begin{equation}
\displaystyle
\widehat{L}_{\s A}=\frac{p_x}{m}\frac{\partial}{\partial x}+\frac{1}{m}\biggl(p_y-\frac{eB}{c}x\biggr)
\biggl(\frac{\partial}{\partial y}+\frac{eB}{c}\frac{\partial}{\partial p_x}\biggr)+
\frac{p_z}{m}\frac{\partial}{\partial z}
\label{lioint}
\end{equation}
%%%
If we compare this Liouville operator with the one containing no interaction with the magnetic field which is
%%%
\begin{equation}
\displaystyle
\widehat{L}=\frac{p_x}{m}\frac{\partial}{\partial x}+\frac{p_y}{m}\frac{\partial}{\partial y}
+\frac{p_z}{m}\frac{\partial}{\partial z} \label{liofree}
\end{equation}
%%%
we see that the tricks to pass from (\ref{liofree}) to (\ref{lioint}) are the substitutions:
%%%
\begin{equation}
\label{classmc}
\left\{
	\begin{array}{l}
	\displaystyle
	p_y\;\longrightarrow\;p_y-\frac{eB}{c}x\smallskip\\
	\displaystyle \frac{\partial}{\partial y}\;\longrightarrow\;\frac{\partial}{\partial y}+\frac{eB}{c}\frac{\partial}
	{\partial p_x}\\
	\end{array}
	\right.
\end{equation}
%%%
These are the MC rules for the Liouville operator in the case of a constant magnetic field.

For the Schr\"odinger operator the MC rules would have been the quantum operatorial version of just the first relation 
of (\ref{classmc})
%%%
\begin{equation}
\displaystyle
\frac{\hbar}{i}\frac{\partial}{\partial y}\;\longrightarrow\;
\frac{\hbar}{i}\frac{\partial}{\partial y}-\frac{eB}{c}x
\end{equation}
%%%
which would turn into the second one if we had represented $\displaystyle x=-\frac{\hbar}{i}
\frac{\partial}{\partial p_x}$. Somehow in QM the two rules (\ref{classmc}) would be just one and the same differently 
than in CM where we cannot identify $p_y$ with $\displaystyle \frac{\partial}{\partial y}$.
\par
In {\bf section 3} of this paper we will generalize the MC rule (\ref{classmc}) to the case of an arbitrary magnetic field. 
Those  rules, even if derived from the same simple principle as above, will   involve various complicated 
combinations of the derivative operators. We shall show anyhow that those complicated combinations could be put in a very simple
and illuminating form using the concept of superfield which naturally appears in a functional approach 
\cite{Gozzi} to the
KvN theory. This functional approach will be  briefly reviewed in {\bf section 2}. 

Let us now go back to our derivation of the minimal coupling rules for $\widehat{L}$. If in the original $H$ there were a
potential $V(y)$ then the $\widehat{L}$ of eq. (\ref{liuvillian}) would have contained a derivative with respect to
$p_y$. Now in (\ref{pipsilon}) we changed $p_y$ to get the MC and so one would be led to conclude that also 
$\displaystyle \frac{\partial}{\partial p_y}$ has to be changed in $\widehat{L}$ and not just
$\displaystyle \frac{\partial}{\partial y}$ as it appears in (\ref{classmc}). Actually this is not the case . We will see
in {\bf section 3} 
that the derivatives with respect to the momenta never have  to be changed in the MC correspondence. 
\par
The reader, realizing that we have an Hilbert space, may wonder if the gauge invariance manifests itself on the states via
a phase like it does in QM. This is so only in a particular representation. The issue of gauge invariance will be 
throughly examined in {\bf section 3} of this paper. 
\par In {\bf section 4} we will apply the KvN theory to the Landau problem. We 
will compare the results to the quantum case and show that there are many more degeneracies at the classical level than 
at the quantum one. Finally in {\bf section 5} we will study the Aharonov-Bohm phenomenon both at the CM level via
the KvN formalism and at the QM level via the Schr\"odinger equation.
We will show that at the {\it quantum} level the spectrum of the Schr\"odinger Hamiltonian is changed by the presence of the 
gauge potential while the spectrum of the {\it classical} Liouville operator is left unchanged. This spectrum 
drives the  motion of the system at the classical level as it is explained in ref. \cite{Mauro}.
Further calculational details omitted in the various sections of this paper are
confined to few appendices which 
conclude the paper.

\section{Functional Approach to the KvN Theory}

As we indicated in the previous section, KvN postulated for the Hilbert space states $\psi$ of CM  
the same evolution (\ref{lio2}) as for  the probability densities $\rho$ (\ref{lio}). So the propagation kernel 
for $\psi$ will be the same as the propagation kernel for $\rho$. This last one has an immediate physical meaning
being the transition probability $P(\varphi^ at|\varphi^ a_0t_0)$ of finding the particle in the phase space point 
$\varphi^a=(q,p)$ at time $t$ if it was in configuration $\varphi^ a_0$ at time $t_0$.
In CM this $P(\varphi^ at|\varphi^ a_{0}t_0)$ is nothing else than a Dirac delta
%%%
\begin{equation}
P(\varphi^ at|\varphi^ a_{0}t_0)=\delta[\varphi^a-\phi^a_{cl}(t;\varphi_0 t_0)]
\end{equation}
%%%
where $\phi^a_{cl}$ is the classical solution of the Hamilton equations of motion $\displaystyle \dot{\varphi}^a=
\omega^{ab}\frac{\partial H}{\partial \varphi^b}$ ($\omega^{ab}$ symplectic matrix) with initial condition 
$(\varphi_{0},t_{0})$. We know that in general, if we have a probability $P(f|i)$ to go from configuration $(i)$ to
configuration $(f)$, the following decomposition holds:
%%%
\begin{equation}
P(f|i)=\sum_{k_i}P(f|k_{\s N-1})P(k_{\s N-1}|k_{\s N-2})\cdots P(k_{\s 1}|1)
\end{equation}
%%%
In our case this becomes
%%%
\begin{equation}
P(\varphi^a t|\varphi^ a_{0}t_0)=\lim_{N\to\infty}\prod_{j=1}^N\int d^{2n}\varphi_{i}\delta^{(2n)}[\varphi^a_j
-\phi^a_{cl}(t_j|\varphi_{j-1}t_{j-1})]
\end{equation}
%%%
where we have sliced the interval of time $t-t_0$ in $N$ intervals labelled by $t_j$.
In the continuum limit we could formally write the relation above as 
%%%
\begin{equation}
P(\varphi^a t|\varphi^ a_{0}t_0)=\int {\cal D}\varphi\;\widetilde{\delta}[\varphi^a(t)-\phi^a_{cl}(t)] \label{prob}
\end{equation}
%%%
where ${\cal D}\varphi$ is a functional integration and $\widetilde{\delta}[\;]$ a functional Dirac delta. As the  
$\phi^a_{cl}(t)$ are the solutions of the Hamilton equation, we could rewrite the $\widetilde{\delta}[\;]$ in (\ref{prob}) as
%%%
\begin{equation}
\widetilde{\delta}[\varphi^a(t)-\phi^a_{cl}(t)]=\widetilde{\delta}[\dot{\varphi}^a-\omega^{ab}\partial_bH]
det[\delta^a_b\partial_t-\partial_b(\omega^{ad}\partial_dH)] \label{prob2}
\end{equation}
%%%
Next we could Fourier transform the Dirac delta on the RHS of (\ref{prob2}) introducing $2n$ extra variables
$\lambda_a$, and we could exponentiate the determinant on the RHS of (\ref{prob2}) using $4n$ anticommuting variables
$c^a,\bar{c}_a$. The final result is the following
%%%
\begin{equation}
\displaystyle
P(\varphi^a t|\varphi^ a_{0}t_0)=\int{\cal D}^{\prime\prime}\varphi{\cal D}\lambda{\cal D}c{\cal D}\bar{c}\,
exp\biggl[i\int_{t_0}^tdt\,\LT\biggr] \label{prob3}
\end{equation}
%%%
where
%%%
\begin{equation}
\LT=\lambda_a\dot{\varphi}^a+i\bar{c}_a\dot{c}^a-\lambda_a\omega^{ab}\partial_bH-i\bar{c}_a\omega^{ad}
(\partial_d\partial_bH)c^b \label{suplag}
\end{equation}
%%%
and with ${\cal D}^{\prime\prime}$ we indicate that the integration is over paths with  fixed end points
in $\varphi$. All this is described in
many more details in  ref. \cite{Gozzi}. Associated to this $\LT$ there is an Hamiltonian which is 
%%%
\begin{equation}
\HT=\lambda_a\omega^{ab}\partial_bH+i\bar{c}_a\omega^{ad}(\partial_d\partial_bH)c^b \label{supham}
\end{equation}
%%%
We notice that, instead of just the original $2n$ phase space coordinates $\varphi^a$, we now have $8n$ variables 
$(\varphi^a,\lambda_a,
c^a,\bar{c}_a)$ whose geometrical meaning has been studied in details in ref. \cite{Regini}.
We will indicate with ${\cal M}$ the original phase space coordinatized by $\varphi^a$ and with $\widetilde{\cal M}$
the space coordinatized by
$(\varphi^a,\lambda_a,c^a,\bar{c}_a)$. What is nice is that all these $8n$ variables can be put 
together in a single object known as superfield in the literature on supersymmetry. In order to construct it we first 
enlarge the standard time $t$ to two Grassmannian
partners $\theta,\bar{\theta}$ which make, together with the time $t$, what is known as superspace and then 
we build the following object
%%%
\begin{equation}
\Phi^a(t,\theta,\bar{\theta})=\varphi^a(t)+\theta
c^a(t)+\bar{\theta}\omega^{ab}\bar{c}_b(t)+i\bar{\theta}\theta\omega^{ab}\lambda_b(t)
\label{super1}
\end{equation}
%%%
It is easy to prove that
%%%
\begin{equation}
i\int d\theta d\bar{\theta} H[\Phi]=\HT \label{superH}
\end{equation}
%%%
where $H$ is the usual Hamiltonian of classical mechanics in ${\cal M}$ but where we have replaced the standard bosonic
variables $\varphi^a$ with the superfield variables $\Phi^a$. By expanding $H(\Phi)$ in $\theta$ and 
$\bar{\theta}$ it is straightforward to prove the following formula
%%%
\begin{equation}
H(\Phi^a)=H(\varphi^a)+\theta N_{\s H}-\bar{\theta}\bar{N}_{\s H}+i\theta\bar{\theta}\HT
\end{equation}
%%%
where the precise form of $N_{\s H},\bar{N}_{\s H}$ is not necessary in this paper and can be found in ref. \cite{Gozzi}.
The same steps  we did above for the Hamiltonian can be done for any function $O(\varphi)$ of the phase space ${\cal M}$,
 i.e. replace $\varphi$ with $\Phi$
and expand $O(\Phi)$ in $\theta,\bar{\theta}$
%%%
\begin{equation}
O(\Phi^a)=O(\varphi^a)+\theta N_{\s O}-\bar{\theta}\bar{N}_{\s O}+i\theta\bar{\theta}{\cal O} \label{supero}
\end{equation}
%%%
where ${\cal O}=\lambda_a\omega^{ab}\partial_bO+i{\bar c}_{a}\omega^{ad}(\partial_{d}\partial_{b}O)c^{b}$. 
The index $"a"$ in $\varphi^a$ indicates either the 
first $n$ configurational variables $q$ or the second $n$ momentum variables $p$, so in the case of $n=1$ the substitution 
in $H$ on the LHS of  (\ref{superH}) is:
%%%
\begin{eqnarray}
&& q\;\longrightarrow \;\Phi^q=q+\theta c^q+\bar{\theta}\bar{c}_p+i\bar{\theta}\theta\lambda_p \label{sup1}\\ 
&& p\;\longrightarrow \;\Phi^p=p+\theta c^p-\bar{\theta}\bar{c}_q-i\bar{\theta}\theta\lambda_q \label{sup2}
\end{eqnarray}
%%%
We have put the index $q$ (or $p$) on $c,\bar{c},\lambda$ just to indicate that we refer to the first $n$ (or the second
$n$) of the $c,\bar{c},\lambda$ variables.
A relation analog to (\ref{superH}) holds also at the Lagrangian level
%%%
\begin{equation}
i\int d\theta d\bar{\theta} L[\Phi]=\LT+(s.t.)
\end{equation}
%%%
where $(s.t.)$ is a surface term of the form
%%%
\begin{equation}
(s.t.)=-\frac{d}{dt}(\lambda_pp+i\bar{c}_pc^p)
\end{equation}
%%%

In this paper we will be interested only in the non-Grassmannian set of variables among the $8n$-ones
$(\varphi^a,\lambda_a, c^a,\bar{c}_a)$, that means only in $\varphi^a,\lambda_a$.
So the $\LT$ and the $\HT$ of formulas (\ref{suplag}) and (\ref{supham}) will be  reduced to\footnote{The subscript $B$ is for
Bosonic but we will neglect it from now on.}:
%%%
\begin{eqnarray}
&&\LT_{\s B}=\lambda_a\dot{\varphi}^a-\lambda_a\omega^{ab}\partial_bH\label{lagbos}\\
&&\HT_{\s B}=\lambda_a\omega^{ab}\partial_bH \label{hambos}
\end{eqnarray}
%%%
and the superfields (\ref{sup1}) and (\ref{sup2}) to:
%%%
\begin{eqnarray}
&&\Phi^q=q+i\bar{\theta}\theta\lambda_p \label{supb1}\\
&&\Phi^p=p-i\bar{\theta}\theta\lambda_q \label{supb2}
\end{eqnarray}
%%%
KvN {\it postulated} for the wave functions $\psi$ the same evolution as for the $\rho$, so the kernel of propagation
for the $\psi$: $K(\varphi^at|\varphi^a_0 t_0)$, also known as transition amplitude,
will have the same path integral expression as the transition
probability $P(\varphi^at|\varphi^a_0 t_0)$. This last has the expression (\ref{prob3}) where actually the integration
over the $c^a,\bar{c}_a$ can be dropped because we are not propagating these variables. Their overall integration
would give one as proved in ref. \cite{Gozzi}. So the equality of $P(\varphi^at|\varphi^a_0 t_0)$
and $K(\varphi^at|\varphi^a_0 t_0)$ would\footnote{The reader may 
be puzzled that the same kernel propagates both $\psi$ and $|\psi|^2$.
This is not a problem and does not lead to any contradiction as proved in appendix A of ref. \cite{Mauro}.} give:
%%%
\begin{equation}
\displaystyle
P(\varphi^at|\varphi^a_0 t_0)=
K(\varphi^a t|\varphi^a_0 t_0)=\int{\cal D}^{\prime\prime}\varphi{\cal D}\lambda \;exp\biggl[ i\int
dt\, \LT_{\s B}\biggr]
\label{kernelpsi}
\end{equation}
%%%
The two path integrals for $P(\;|\;)$ and for $K(\;|\;)$ are the same if we want to propagate probability densities
$\rho$ or wave functions $\psi$ both living in the $\varphi$-space only.
If instead we want to propagate the analogous objects living in the
$(\varphi,c)$-space, then the evolution of $\rho(\varphi,c)$ would be via the $\LT$ of eq. (\ref{suplag}) but 
the evolution of the $\psi(\varphi,c)$ would be via a different 
Lagrangian studied in ref. \cite{metaplectic}.
\par 
In a manner similar to what is done in field theory we could also introduce the
generating functional which would have the form
%%%
\begin{equation}
\displaystyle
Z_{\s CM}[j]=\int {\cal D}\varphi{\cal D}\lambda exp\biggl[i\int dt\,[\LT_{\s B}+j_{\varphi}\varphi+j_{\lambda}\lambda]\biggr]
\end{equation}
%%%
where $j_{\varphi},j_{\lambda}$ are currents associated to the two variables $\varphi,\lambda$.
We shall now derive the operatorial formalism associated to this path integral and compare it with the KvN operatorial
version of CM given by eq. (\ref{lio2}). Having a path integral we can introduce the concept of commutator as Feynman did in
the quantum case: given two functions $O_1(\varphi,\lambda)$ and $O_2(\varphi,\lambda)$, let us  evaluate the following
quantity under our path integral:
%%%
\begin{equation}
\langle[O_1,O_2]\rangle\equiv\lim_{\epsilon\to 0}\langle O_1(t+\epsilon)O_2(t)-O_2(t+\epsilon)O_1(t)\rangle
\end{equation}
%%%
which is defined  as the expectation value of the commutator. In our case what we get is
%%%
\begin{eqnarray}
&&\langle[\varphi^a,\varphi^b]\rangle=0 \label{comm0}\\
&&\langle[\varphi^a,\lambda_b]\rangle=i\delta_b^a \label{comm}
\end{eqnarray}
%%%
The first commutators confirm that we are doing CM and not QM. In fact they tell us that the $q$ commute not only among
themselves but also with the $p$. The second commutators instead tell us that the $\lambda_a$ are something like the momenta
conjugate to $\varphi^a$. In order to satisfy the (\ref{comm}) they can be realized as
%%%
\begin{equation}
\displaystyle
\widehat{\varphi}^a=\varphi^a, \;\;\;\;\widehat{\lambda}_a=-i\frac{\partial}{\partial\varphi^a} \label{oprep}
\end{equation}
%%%
Via this operatorial realization of $\lambda_a$ also $\HT$ of eq. (\ref{hambos}) can be turned into an operator
%%%
\begin{equation}
\HT_{\s B}\;\longrightarrow\;\HCT=-i\omega^{ab}\partial_bH\partial_a \label{operat}
\end{equation}
%%%
and it is easy to see that $\HCT
=-i\widehat{L}$ where $\widehat{L}$ is the Liouville operator of eq.
(\ref{liuvillian}). So this confirms that the operatorial formalism generated by our path integral is nothing else than the KvN
one.

If we specify things to $n=1$, the relations (\ref{comm})
are just  $2$:
%%%
\begin{eqnarray}
&&[q,\lambda_q]=i\label{comm1}\\
&&[p,\lambda_p]=i\label{comm2}
\end {eqnarray}
%%%
and the representation we used for $\lambda_q$ in (\ref{oprep}) is not the only possible one.
In fact in  (\ref{oprep}) we realized $\varphi^a$ as a multiplicative operator and $\lambda_a$ as a derivative one but, using
the fact that (\ref{comm1}) and (\ref{comm2}) are two distinct relations, we could have realized $q$ and $\lambda_p$ as
multiplicative operators and $\lambda_q$ and $p$ as derivative ones:
%%%
\begin{equation}
\left\{
	\begin{array}{l}
	\displaystyle q \longrightarrow \widehat{q}\smallskip\\
          \displaystyle \lambda_p \longrightarrow \widehat{\lambda}_p\smallskip\\
          \displaystyle \lambda_q  \longrightarrow -i\frac{\partial}{\partial q}\smallskip\\
          \displaystyle p \longrightarrow i\frac{\partial}{\partial \lambda_p} \label{mixrep}
	\end{array}
	\right.
\end{equation}
%%%
In the representation (\ref{oprep}) we would  diagonalize $\widehat{\varphi}^a$ 
%%%
\begin{equation}
\widehat{\varphi}^a|q,p\rangle=\varphi^a|q,p\rangle \label{diagmix0}
\end{equation}
%%%
and the kernel (\ref{kernelpsi}) could be formally represented as
%%%
\begin{equation}
\label{mario}
K(\varphi^a t|\varphi^a_0 t_0)=\langle qpt|q_0p_0t_0\rangle
\end{equation}
%%%
In the mixed representation (\ref{mixrep}) we would diagonalize instead these other operators:
%%%
\begin{eqnarray}
&&\widehat{q}|q,\lambda_p\rangle=q|q,\lambda_p\rangle\nonumber\\
&&\widehat{\lambda}_p|q,\lambda_p\rangle=\lambda_p|q,\lambda_p\rangle \label{diagmix}
\end{eqnarray}
%%%
In this  representation we have to do a "partial Fourier transform" in order to go from the states
$|qp\rangle$ to the $|q\lambda_p\rangle$ ones . We say "partial Fourier transform" because somehow we are just "replacing"
$p$ in the state $|q p\rangle$ with $\lambda_p$ which is somehow the momentum conjugate to $p$ as can be
easily seen from the Lagrangian $\LT_{\s B}$ of eq. (\ref{lagbos}). Using the transformation formula\footnote{This is the
"analog" of the transformation formula $\displaystyle \langle q|p\rangle=\frac{1}{\sqrt{2\pi\hbar}}exp[ipq/\hbar]$ between
momenta and position eigenstates but here the role of position and momenta is taken respectively by $p$ and $\lambda_p$.}
%%%
\begin{equation}
\displaystyle
\langle
q^{\prime}p^{\prime}|q,\lambda_p\rangle=\frac{1}{\sqrt{2\pi}}\delta(q-q^{\prime})exp\bigl[ip^{\prime}\lambda_p\bigr]
\label{fourier}
\end{equation}
%%%
we can transform $\psi$ from the $|q,p\rangle$ basis to the $|q,\lambda_p\rangle$ one:
%%%
\begin{equation}
{\widetilde{\psi}}(q,\lambda_p)\equiv\langle q,\lambda_p|\psi\rangle=\int dq^{\prime}dp^{\prime}
\langle q\lambda_p|q^{\prime}p^{\prime}\rangle\cdot
\langle q^{\prime}p^{\prime}|\psi\rangle=\frac{1}{\sqrt{2\pi}}\int \psi(q,p)e^{-i\lambda_pp}dp
\end{equation}
%%%
In the same manner we can transform the kernel (\ref{mario}) and get:
%%%
\begin{eqnarray}
\displaystyle
&&{\widetilde{K}}(q\lambda_pt|q_0\lambda_{p_0}t_0)\equiv\langle q\lambda_p t|q_0
\lambda_{p_0}t_0\rangle=\nonumber\\
&&=\int dq^{\prime}_0dp^{\prime}_0 dq^{\prime}dp^{\prime}\langle
q\lambda_pt|q^{\prime}p^{\prime}t\rangle\langle q^{\prime}p^{\prime}t|
q^{\prime}_0p^{\prime}_0t_0\rangle\langle q_0^{\prime}p_0^{\prime}t_0|q_0\lambda_{p_0}t_0\rangle=\nonumber\\
&&=\int dq_0^{\prime}dp_0^{\prime}dq^{\prime}dp^{\prime}\frac{1}{\sqrt{2\pi}}e^{-i\lambda_pp^{\prime}}\delta(q-q^{\prime})
K(q^{\prime}p^{\prime}t^{\prime}|q^{\prime}_0p^{\prime}_0t_0)\frac{1}{\sqrt{2\pi}}e^{i\lambda_{p_0}p_0^{\prime}}
\delta(q_0^{\prime}-q_0)=\nonumber\\
&&=\frac{1}{2\pi}\int
dp_0^{\prime}dp^{\prime}exp[-i\lambda_pp^{\prime}]exp[i\lambda_{p_0}p_0^{\prime}] K(qp^{\prime}t|q_0p_0^{\prime}t_0)
\end{eqnarray}
%%%
This mixed representation is the most useful one in analyzing the issue of how to go from this formulation of CM to
QM \cite{Abrikosov}. It is also the representation where gauge invariance manifests itself via the multiplication by a phase
like in QM. We will examine all this in details in the next section.

\section{Minimal Coupling in $\HT$ and Gauge Invariance}

In the previous section we have introduced the concept of superfield, formulas
(\ref{super1})-(\ref{sup1})-(\ref{sup2})-(\ref{supb1})-(\ref{supb2}). Using them it is 
then easy to put in a compact form the
minimal coupling rules (\ref{classmc}) used in the case of a constant magnetic 
field. First, via the representation
(\ref{oprep}), the relation (\ref{classmc}) can be turned into the following  one
%%%
\begin{equation}
\label{classmc2}
\left\{
	\begin{array}{l}
	\displaystyle p_y\;\longrightarrow\;p_y-\frac{eB}{c}x\smallskip\\
          \displaystyle \lambda_y\;\longrightarrow\;\lambda_y+\frac{eB}{c}\lambda_{p_x}\\
	\end{array}
	\right.
\end{equation}
%%%
Next let us notice what we did in (\ref{super1})-(\ref{superH}) in order to go from the Hamiltonian $(H)$ in
${\cal M}$ to the new Hamiltonian $(\HT)$ in $\widetilde{\cal M}$: we just replaced $\varphi^a$ with the superfield $\Phi^a$.
Let us do the same for the minimal coupling in ${\cal M}$ space 
for the case of a constant
magnetic field given by eq. (\ref{pipsilon}). 
It means the following:
%%%
\begin{eqnarray}
\displaystyle
p_y\;&\longrightarrow &\;p_y-\frac{e}{c}Bx \label{seia}\\
\downarrow \;& &\;\;\;\;\;\;\;\;\downarrow\nonumber\\
\Phi^{p_y}\;&\longrightarrow &\Phi^{p_y}-\frac{e}{c}B\Phi^{x} \label{seib}
\end{eqnarray}
%%%
Expanding (\ref{seib}) in $\theta,\bar{\theta}$ and using (\ref{supb1}) and (\ref{supb2}) we get 
%%%
\begin{equation}
\displaystyle
p_y-i\bar{\theta}\theta\lambda_y \;\longrightarrow\; p_y-i\bar{\theta}\theta \lambda_y-\frac{eB}{c}(x+i\bar{\theta}
\theta\lambda_{p_x}) \label{sette}
\end{equation}
%%%
Comparing the terms with equal number of $\theta$ and $\bar{\theta}$ (\ref{sette}) becomes
%%%
\begin{equation}
\label{otto}
\left\{
	\begin{array}{l}
	\displaystyle p_y\;\longrightarrow\;p_y-\frac{eB}{c}x\smallskip\\
          \displaystyle \lambda_y\;\longrightarrow\;\lambda_y+\frac{eB}{c}\lambda_{p_x}\\
	\end{array}
	\right.
\end{equation}
%%%
which are exactly the substitution rules (\ref{classmc2}) for the minimal coupling for $\HT$. So we can say that the
superfield formalism provides a compact way, eq. (\ref{seib}), to write the complicated minimal coupling (\ref{classmc2}). 

Let us now check if this compact way of expressing things via superfields is an accident of the case of a constant magnetic field or
if it holds in general. The Hamiltonian $H$ of a free particle in a generic magnetic field is obtained via the minimal coupling (\ref{mincoup}) and is
%%%
\begin{equation}
\displaystyle
H=\frac{1}{2m}\biggl\{\biggl(p_x-\frac{e}{c}A_x\biggr)^2+\biggl(p_y-\frac{e}{c}A_y\biggr)^2+
\biggl(p_z-\frac{e}{c}A_z\biggr)^2\biggr\} \label{hmagn}
\end{equation}
%%%
The associated $\HT_{\s B}$ of (\ref{hambos}) is then:
%%%
\begin{eqnarray}
\displaystyle
&&\HT_{\s B}=\frac{\lambda_x}{m}\biggl(p_x\,-\frac{e}{c}A_{x}\biggr)+\frac{\lambda_y}{m}\biggl(p_y-\frac{e}{c}A_{y}\biggr)
+\frac{\lambda_z}{m}\biggl(p_z-\frac{e}{c}A_{z}\biggr)\nonumber\\
&&\qquad\qquad\qquad-\lambda_{p_x}\frac{\partial H}{\partial x}-
\lambda_{p_y}\frac{\partial H}{\partial y}-\lambda_{p_z}\frac{\partial H}{\partial z}=\nonumber\\
&&=\frac{1}{m}\biggl(\lambda_x+\frac{e}{c}\frac{\partial A_x}{\partial x}\lambda_{p_x}+\frac{e}{c}\frac{\partial
A_x}{\partial y}\lambda_{p_y}+\frac{e}{c}\frac{\partial A_x}{\partial
z}\lambda_{p_z}\biggr)\biggl(p_x-\frac{e}{c}A_x\biggr)\nonumber\\
&&\;\,+\frac{1}{m}\biggl(\lambda_y+\frac{e}{c}\frac{\partial A_y}{\partial x}\lambda_{p_x}+\frac{e}{c}\frac{\partial
A_y}{\partial y}\lambda_{p_y}+\frac{e}{c}\frac{\partial A_y}{\partial
z}\lambda_{p_z}\biggr)\biggl(p_y-\frac{e}{c}A_y\biggr)\nonumber\\
&&\;\,+\frac{1}{m}\biggl(\lambda_z+\frac{e}{c}\frac{\partial A_z}{\partial x}\lambda_{p_x}+\frac{e}{c}\frac{\partial
A_z}{\partial y}\lambda_{p_y}+\frac{e}{c}\frac{\partial A_z}{\partial
z}\lambda_{p_z}\biggr)\biggl(p_z-\frac{e}{c}A_z\biggr) \label{hbosmag}
\end{eqnarray}
%%%
So we see that this last expression can be obtained from the $\HT$ of the free particle
%%%
\begin{equation}
\displaystyle
\HT=\frac{1}{m}\lambda_xp_x+\frac{1}{m}\lambda_yp_y+\frac{1}{m}\lambda_zp_z
\end{equation}
%%%
via the simple substitution
%%%
\begin{equation}
\label{general}
\displaystyle
\left\{
	\begin{array}{l}
	\displaystyle p_i\;\longrightarrow\;p_i-\frac{e}{c}A_{q_i}\smallskip\\
          \displaystyle \lambda_{q_i}\;\longrightarrow\;\lambda_{q_i}+\frac{e}{c}\sum_j\biggl(\lambda_{p_j}
          \frac{\partial A_{q_i}}{\partial q_j}\biggr)\\
	\end{array}
	\right.
\end{equation}
%%%
This is the minimal coupling for $\HT$ in a generic magnetic field generalizing the one in a 
constant magnetic field given by eq. 
(\ref{otto}). We have indicated with $A_{q_i}$ the spatial components of the gauge field. We want now to see if 
(\ref{general}) can  be derived from the superfield generalization of the standard MC, i.e.:
%%%
\begin{eqnarray}
\displaystyle
p_i\;&\longrightarrow &\;p_i-\frac{e}{c}A_{q_i}(q) \label{duea}\\
\downarrow \;& &\;\;\;\;\;\;\;\;\downarrow\nonumber\\
\Phi^{p_i}\;&\longrightarrow &\Phi^{p_i}-\frac{e}{c}A_{q_i}(\Phi^q)\label{dueb}
\end{eqnarray}
%%%
Let us first notice that, if we neglect the $c,\bar{c}$ variables, we have that the following relation holds:
%%%
\begin{equation}
\displaystyle
i\int d\theta d\bar{\theta} A_{q_i}[\Phi^q]=-\sum_j\lambda_{p_j}\frac{\partial A_{q_i}}{\partial q_j}\equiv
{\mathcal A}_{q_i} \label{tre}
\end{equation}
%%%
This  can be checked easily
using eq. (\ref{supero}).
If we now expand  (\ref{dueb}) in $\theta,{\bar\theta}$ and we neglect the $c,\bar{c}$, what we  get is:
%%%
\begin{equation}
p_i-i\bar{\theta}\theta\lambda_{q_i}
\;\longrightarrow\;p_i-i\bar{\theta}\theta\lambda_{q_i}-\frac{e}{c}(A_{q_i}+i\theta\bar{\theta}
{\mathcal A}_{q_i})
\end{equation}
%%%
Comparing the terms above with the same number of $\theta,\bar{\theta}$ we obtain
%%%
\begin{equation}
\displaystyle
\left\{
	\begin{array}{l}
	\displaystyle p_i\;\longrightarrow\;p_i-\frac{e}{c}A_{q_i}\smallskip\\
 \displaystyle \lambda_{q_i}\;\longrightarrow\;\lambda_{q_i}+\frac{e}{c}\sum_j\biggl(\lambda_{p_j}
          \frac{\partial A_{q_i}}{\partial q_j}\biggr)\\
	\end{array}
	\right.
\end{equation}
%%%
which are exactly the relations (\ref{general}). So this proves that (\ref{dueb}) is the most compact way to write the minimal coupling
for $\HT$.

The reader may ask which is the physical reason why $\lambda_{q_i}$ should be changed, like we do in eq. (\ref{general}),
when we turn on the magnetic field. To answer this question we have to analyze the issue of the gauge invariance
of the system. Let us remember that the Lagrangian associated to the $H$ of eq. (\ref{hmagn}) was
%%%
\begin{equation}
L=\frac{1}{2}m(\dot{x}^2+\dot{y}^2+\dot{z}^2)+\frac{e}{c}(\dot{x}A_x+\dot{y}A_y+\dot{z}A_z) \label{lmagn}
\end{equation}
%%%
where
%%%
\begin{equation}
\displaystyle
\label{velocities}
\left\{
	\begin{array}{l}
	\displaystyle \dot{x}=\frac{1}{m}\bigl(p_x-\frac{e}{c}A_x\bigr)\smallskip\\
          \displaystyle \dot{y}=\frac{1}{m}\bigl(p_y-\frac{e}{c}A_y\bigr)\smallskip\\
          \displaystyle \dot{z}=\frac{1}{m}\bigl(p_z-\frac{e}{c}A_z\bigr)\smallskip\\
	\end{array}
	\right.
\end{equation}
%%%
The velocities which appear above are measurable quantities and so they must be gauge-invariant. Since 
$A_{q_i}$ transform under a gauge
transformation as $A_{q_i}+\partial_{q_i}\alpha(q)$ with $\alpha(q)$ an arbitrary function, the $p_i$
must transform as
%%%
\begin{equation} 
p_i\;\longrightarrow\;p_i+\frac{e}{c}\partial_{q_i}\alpha(q) \label{six}
\end{equation}
%%%
The Hamiltonian $H$, being a combination of gauge-invariant quantities like $\displaystyle p_i-\frac{e}{c}A_{q_i}$, 
is gauge invariant while the
$L$ of eq. (\ref{lmagn}) changes by a total derivative. As the Hamiltonian $H$ is basically the energy of the system
it must be gauge invariant. Let us now ask ourselves how $\lambda$ should change under a gauge transformation. If we adopt the
compact notation (\ref{dueb}) we used for the minimal coupling of $\HT$, it seems natural that the superfield analog of the gauge
transformation (\ref{six}) should be
%%%
\begin{equation}
\Phi^{p_i}\;\longrightarrow\;\Phi^{p_i}+\frac{e}{c}[\partial_{q_i}\alpha](\Phi^{q}) \label{sevena}
\end{equation}
%%%
and
%%%
\begin{equation}
A_{q_i}(\Phi^q)\;\longrightarrow\;A_{q_i}(\Phi^q)+[\partial_{q_i}\alpha](\Phi^q) \label{sevenb}
\end{equation}
%%%
where by $[\partial_{q_i}\alpha](\Phi^q)$ we mean that we insert $\Phi^q$ in place of $q$ in the function
that we obtain by making the derivative of $\alpha(q)$ with respect to $q_i$.
Expanding (\ref{sevena}) in $\theta,\bar{\theta}$ we get (\ref{six}) as first component and the following one as last
component
%%%
\begin{equation}
\lambda_{q_i}\;\longrightarrow\;\lambda_{q_i}+i\int d\theta d\bar{\theta}\;\frac{e}{c}[\partial_{q_i}\alpha](\Phi^q)
=\lambda_{q_i}-\frac{e}{c}\lambda_{p_j}\partial_{q_j}\partial_{q_i}\alpha(q) \label{lamtra}
\end{equation}
%%%
Similarly expanding (\ref{lamtra}) in $\theta, \bar{\theta}$ we get the usual transformation 
$A_{q_i}\longrightarrow A_{q_i}+\partial_{q_i}\alpha$ as first component and the following one as last component:
%%%
\begin{equation}
{\mathcal A}^{\prime}_{q_i}={\mathcal A}_{q_i}+\partial_{q_i}\widetilde{\alpha}(q,\lambda_p)
\end{equation}
%%%
where  
%%%
\begin{equation}
\displaystyle \widetilde{\alpha}(q,\lambda_p)=-\sum_j\lambda_{p_j}\frac{\partial\alpha}{\partial q_j}
\label{aexp}
\end{equation}
%%% 
It is then easy to see that the combination $\displaystyle \lambda_{q_i}-\frac{ie}{c}\int d\theta d\bar{\theta}
A_{q_i}(\Phi^q)\equiv\lambda_{q_i}-\frac{e}{c}{\mathcal A}_{q_i}$, which is the $\lambda_{q_i}$ components 
of eq. (\ref{dueb}), is gauge invariant if we gauge transform $\lambda_{q_i}$ as in (\ref{lamtra}) and $A_{q_i}(\Phi^q)$ as in 
(\ref{sevenb}). So the RHS of (\ref{general}) are gauge invariant quantities and, as a consequence, also the $\HT_{\s B}$ 
of eq. (\ref{hbosmag}) is gauge invariant because it is built out of the combinations (\ref{general}). 
Of course all this is very formal and stems
from the extension of the standard gauge transformations to the superfields, (\ref{sevena})-(\ref{sevenb}). 
While $H$, being the energy, must be gauge invariant there is apparently no reason why $\HT$ should be gauge invariant.
Similarly, while we know that $p_i$ should change under a gauge transformation like in (\ref{six}) in order to make
gauge invariant the velocities (\ref{velocities}) which are observables, there is apparently no physical reason why
$\lambda_{q_i}$ should change under a gauge transformation as in (\ref{lamtra}). Actually there is a physical reason 
and it is  the
following. As the velocities (\ref{velocities}) are gauge invariant then their evolution has to be gauge invariant too. 
The evolution can
occur via $H$ and the  standard Poisson brackets
%%%
\begin{equation}
\{\varphi^a,\varphi^b\}_{pb}=\omega^{ab}
\end{equation}
%%%
or via $\HT$ and some extended Poisson brackets $\{\;,\;\}_{epb}$ which were introduced in ref. \cite{Gozzi} (for a
brief review see appendix A). In this extended formalism the Hamiltonian $\HT$ (\ref{hbosmag}) can be written,
using the notation
$(\ref{tre})$, in the following compact way:
%%%
\begin{equation}
\displaystyle
\HT=\frac{1}{m}\sum_i\biggl(\lambda_{q_i}-\frac{e}{c}{\mathcal A}_{q_i}\biggr)\biggl(p_i-\frac{e}{c}A_{q_i}\biggr)
\label{abstham}
\end{equation}
%%%
The gauge invariant velocities
%%%
\begin{equation}
v_{q_i}=\frac{1}{m}\biggl(p_i-\frac{e}{c}A_{q_i}\biggr)
\end{equation}
%%%
evolve via the extended Poisson brackets in the following manner: 
%%%
\begin{equation}
\dot{v}_{q_i}=\{v_{q_i},\HT\}_{epb}
\end{equation}
%%%
For example the result  for $v_x$ is
%%%
\begin{equation}
\displaystyle
\dot{v}_x=\frac{e}{mc}(B_zv_y-B_yv_z)
\end{equation}
%%%
If we now use a different gauge and just change the gauge field $A_{q_i}$ and $p_i$, {\it but not} 
$\lambda_{q_i}$, we would
get as new $\HT$:
%%%
\begin{equation}
\displaystyle
\HT^{\prime}=\frac{1}{m}\sum_i\biggl(\lambda_{q_i}-\frac{e}{c}{\mathcal A}_{q_i}^{\prime}\biggr)\biggl(p_i^{\prime}-
\frac{e}{c}A^{\prime}_{q_i}\biggr)
\end{equation}
%%%
where with $A^{\prime}_{q_i},p_i^{\prime}$ and ${\mathcal A}^{\prime}_{q_i}$ we indicate the quantities gauge
transformed according to (\ref{sevena})-(\ref{sevenb}).

The evolution
of the velocity $v_x$, via  the gauge transformed $\HT^{\prime}$, would turned out to be:
%%%
\begin{equation}
\displaystyle
\dot{v}_x=\{v_x,\HT^{\prime}\}_{epb}=\frac{e}{mc}(B_zv_y-B_yv_z)+\frac{e}{mc}[(\partial_x^2\alpha)
v_x+(\partial_y\partial_x\alpha)v_y+(\partial_z\partial_x\alpha)v_z] \label{dependence}
\end{equation}
%%%
So we notice that the evolution is not anymore gauge invariant because it depends on the gauge parameters $\alpha$ which
appear on the RHS of (\ref{dependence}). This is absurd because the velocities are gauge invariant quantities and so their
evolution must maintain their gauge invariance. This lack of gauge invariance is the price we would have paid by  not allowing
$\lambda_{q_{i}}$ to change under a gauge transformation or, equivalently, by not allowing $\lambda$ to enter $\HT$ via the MC 
combination of eq. (\ref{general}). Note  that the $\lambda_{p_{i}}$, differently than the
$\lambda_{q_{i}}$, do not have to be changed at all in order to maintain the gauge invariance of the system. This answers
one question we raised in the introduction and concludes the analysis of the issue of the gauge invariance.
  
Up to now we have regarded $\HT$ as a function endowed with its own
extended Poisson brackets. We want now to proceed to analyze the same issue of gauge invariance
when we turn $\HT$ into an operator
$\HCT$ like we did in formula (\ref{operat}). Let us first briefly review what
happens in QM, following ref. \cite{quantumbooks}. In classical mechanics the gauge transformations leave $q$ invariant 
but change $p_i$ as follows
%%%
\begin{equation}
p_i^{\prime}=p_i+\frac{e}{c}\partial_{q_i}\alpha(q) \label{classgau}
\end{equation}
%%%
and, as a consequence, the  Poisson brackets are left invariant under these transformations.
%%%
\begin{equation}
\{q_i,p_j\}=\{q_i,p_j^{\prime}\}=\delta_{ij}
\end{equation}
%%%
Then the quantization rules for the transformed variables are 
%%%
\begin{equation}
\{q_i,p_j^{\prime}\}=\delta_{ij}\;\longrightarrow\;[\widehat{q}_i,\widehat{p}_j^{\,\prime}]=i\hbar\delta_{ij} \label{brack}
\end{equation}
%%%
This implies that $\displaystyle \widehat{p}_j^{\,\prime}$ can be realized operatorially like the original $\widehat{p}_j$, i.e.
 $\displaystyle \widehat{p}_j^{\,\prime}=-i\hbar\frac{\partial}{\partial q_j}$. The quantum Hamiltonian is then 
%%%
\begin{equation}
\displaystyle
\widehat{H}=\biggl(-i\hbar\frac{\partial}{\partial q_i}-\frac{e}{c}\widehat{A}_{q_i}\biggr)^2\bigg/2m
\end{equation}
%%%
As the $\widehat{p}^{\,\prime}_i$ have been realized as before, if we do a gauge transformation  we get that only 
$\widehat{A}_i$
change in ${\widehat H}$ and the new Hamiltonian is:
%%%
\begin{equation}
\displaystyle
\widehat{H}^{\prime}=\biggl[-i\hbar\frac{\partial}{\partial
q_i}-\frac{e}{c}(\widehat{A}_{q_i}+\partial_{q_i}\widehat{\alpha})\biggr]^2\bigg/2m
\end{equation}
%%%
It is easy to check that one can pass from $\widehat{H}$ to $\widehat{H}^{\prime}$ via a unitary transformation
%%%
\begin{equation}
\widehat{H}^{\prime}=U\widehat{H}U^{-1} \label{unit}
\end{equation}
%%%
where $\displaystyle U=exp\biggl(i\frac{e}{c\hbar}\alpha(\widehat{q})\biggr)$. So, differently than 
for the function $H$ of CM, the Hamiltonian is
not gauge invariant, but what is important is that the expectation values are gauge invariant. In fact, if the $H$
transforms as (\ref{unit}), the states change as
%%%
\begin{equation}
|\psi^{\prime}\rangle=U|\psi\rangle
\end{equation}
%%%
In the $\langle q|$ representation this becomes
%%%
\begin{equation}
\displaystyle
\psi^{\prime}(q)=exp\biggl(i\frac{e}{c\hbar}\alpha(q)\biggr)\psi(q)
\end{equation}
%%%
which is the usual transformation by a phase under gauge transformations. We can notice that the expectation values of 
$\langle\psi^{\prime}|\widehat{p}^{\,\prime}_i|\psi^{\prime}\rangle$ and $\langle\psi|\widehat{p}_i|\psi\rangle$ are related
exactly as the classical momenta in (\ref{classgau}):
%%%
\begin{equation}
\langle\psi^{\prime}|\widehat{p}_i^{\,\prime}|\psi^{\prime}\rangle=
\langle\psi|\widehat{p}_i|\psi\rangle+\frac{e}{c}\partial_{q_i}\alpha(q)
\end{equation}
%%%

Let us now turn to the KvN operatorial theory and check how the gauge transformations are implemented. At the
operatorial level we have to construct everything  so that the expectation values 
of
%%%
\begin{equation}
\langle\psi|\widehat{p}_i-\frac{e}{c}\widehat{A}_{q_i}|\psi\rangle\;\;and\;of\;\;
\langle\psi|\widehat{\lambda}_{q_i}-\frac{e}{c}\widehat{{\mathcal A}}_{q_i}|\psi\rangle \label{gaugeinv}
\end{equation}
%%%
would be gauge invariant. Let us start by noticing that the commutation
relations $[\widehat{\varphi}^a,\widehat{\lambda}_b]=i\delta^a_b$ are the operatorial counterpart of
the extended Poisson brackets\break $\{\varphi^a,\lambda_b\}_{epb}=\delta_b^a$ and the  gauge transformed
coordinates  $\varphi^{\prime a},\lambda_b^{\prime}$ under (\ref{six}) and (\ref{lamtra}) have the same epb
$\{\varphi^{\prime a},\lambda^{\prime}_b\}_{epb}=\delta_b^a$ as the original variables. So we expect that 
also the associated commutators among the gauge transformed operators would be the same as the original one:
%%%
\begin{equation}
[\widehat{\varphi}^{\prime a},\widehat{\lambda}^{\prime}_b]=i\delta_b^a
\end{equation}
%%%
This means that we can represent the $\widehat{\varphi}^{\prime a}$ and 
$\widehat{\lambda}^{\prime}_b$ in the same manner as
the $\widehat{\varphi}^a$ and $\widehat{\lambda}_b$. As a consequence  the gauge transformed version of 
the quantities in (\ref{gaugeinv}) is:
%%%
\begin{eqnarray}
&&\langle\psi^{\prime}|\widehat{p}_i-\frac{e}{c}\widehat{A}_{q_i}-\frac{e}{c}[\partial_{q_i}\alpha](\widehat{q})|\psi^{\prime}
\rangle\nonumber\\
&&\langle\psi^{\prime}|\widehat{\lambda}_{q_i}-\frac{e}{c}\widehat{{\mathcal A}}_{q_i}-\frac{e}{c}[\partial_{q_i}
\widetilde{\alpha}](\widehat{q},\widehat{\lambda}_p)|\psi^{\prime}\rangle \label{exp}
\end{eqnarray}
%%%
Note that, via the introduction of the following operator
%%%
\begin{equation}
\widetilde{U}=exp\bigg\{-i\frac{e}{c}\widehat{\lambda}_{p_i}[\partial_{q_i}\alpha](\widehat{q})\biggr\} \label{uexp}
\end{equation}
%%%
we can write the following transformations:
%%%
\begin{eqnarray}
&&\widehat{p}_i-\frac{e}{c}\widehat{A}_{q_i}-\frac{e}{c}[\partial_{q_i}\alpha](\widehat{q})=\widetilde{U}
\biggl[\widehat{p}_i-\frac{e}{c}
\widehat{A}_{q_i}\biggr]\widetilde{U}^{-1}\nonumber\\
&&\widehat{\lambda}_{q_i}-\frac{e}{c}\widehat{{\mathcal A}}_{q_i}-\frac{e}{c}[\partial_{q_i}
\widetilde{\alpha}](\widehat{q},\widehat{\lambda}_p)=
\widetilde{U}\biggl[\widehat{\lambda}_{q_i}-\frac{e}{c}\widehat{{\mathcal A}}_{q_i}\biggr]\widetilde{U}^{-1}
\end{eqnarray}
%%% 
This implies that (\ref{exp}) will be gauge invariant provided we transform the states as follows
%%%
\begin{equation}
|\psi^{\prime}\rangle=\widetilde{U}|\psi\rangle=exp\biggl\{-i\frac{e}{c}\widehat{\lambda}_{p_i}[\partial_{q_i}\alpha]
(\widehat{q})\biggr\}|\psi\rangle \label{abstract}
\end{equation}
%%%
Let us now represent this transformation law on the two basis given by (\ref{diagmix0}) and (\ref{diagmix}). In the basis
(\ref{diagmix}) we have from (\ref{abstract})
%%%
\begin{eqnarray}
\psi^{\prime}(q,\lambda_p)&\equiv&\langle q\lambda_p|\psi^{\prime}\rangle=\langle
q\lambda_p|exp\biggl\{-i\frac{e}{c}\widehat{\lambda}_{p_i}[\partial_{q_i}\alpha](\widehat{q})\biggr\}|\psi\rangle=
\nonumber\\
&=&exp\biggl\{-i\frac{e}{c}\lambda_{p_i}\partial_{q_i}\alpha(q)\biggr\}\langle
q\lambda_p|\psi\rangle=\nonumber\\
&=&exp\biggl(i\frac{e}{c}\widetilde{\alpha}\biggr)\psi(q,\lambda_p)
\label{sixtysix}
\end{eqnarray}
%%%
So in this basis the gauge transformation is just the multiplication by a {\it local phase factor} $\widetilde{\alpha}$
in the space $(q,\lambda_p)$ where $\widetilde{\alpha}$ has been defined in (\ref{aexp}).

Now let us  represent (\ref{abstract}) in the $\langle q,p|$ basis (\ref{diagmix0}) and let us make use 
of the transformation formula (\ref{fourier}). What we get is:
%%%%
\begin{eqnarray}
\displaystyle
\psi^{\prime}(q,p)&=&\langle qp|\psi^{\prime}\rangle=\int dq^{\prime}d\lambda_p^{\prime}\langle
qp|q^{\prime}\lambda_p^{\prime}\rangle\langle q^{\prime}\lambda_p^{\prime}|\widetilde{U}|\psi\rangle \nonumber\\
&=&\int
\frac{d\lambda_p^{\prime}}{\sqrt{2\pi}}exp[ip\lambda_p^{\prime}]exp\biggl[i\frac{e}{c}\widetilde{\alpha}
(q,\lambda_p^{\prime})\biggr]\langle
q\lambda_p^{\prime}|\psi\rangle 
\end{eqnarray}
%%%
Inserting a further completeness we obtain
%%%
\begin{eqnarray}
\displaystyle
\psi^{\prime}(q,p)&=&\int
d\lambda_p^{\prime}dq^{\prime}dp^{\prime}exp[ip\lambda_p^{\prime}]exp\biggl[i\frac{e}{c}\widetilde{\alpha}(q,\lambda_p^{\prime})
\biggr]
\langle q\lambda_p^{\prime}|q^{\prime}p^{\prime}\rangle\langle q^{\prime}p^{\prime}|\psi\rangle=\nonumber\\
&=&\int
\frac{d\lambda_p^{\prime}dp^{\prime}}{2\pi}exp\biggl[i\lambda_p^{\prime}\bigl(p-p^{\prime}-\frac{e}{c}\partial\alpha\bigr)
\biggr]\psi(q,p^{\prime})=\nonumber\\ &=&\int
dp^{\prime}\delta\biggl(p-p^{\prime}-\frac{e}{c}\partial\alpha\biggr)\psi(q,p^{\prime})=
\psi\biggl(q,p-\frac{e}{c}\partial\alpha\biggr)
\end{eqnarray}
%%%
So in the $(q,p)$ representation of our Hilbert space 
the gauge transformations are not implemented by the multiplication by
a local phase, like in the $(q,\lambda_p)$ representation, but by just a shift in the argument $p$ of the wave function. It is
easy to show, as we will do in the appendix B, that the Liouville eq. (\ref{lio2}) is invariant in form under the gauge
transformations. In appendix C we will show that the phase 
$\displaystyle exp\biggl(i\frac{e}{c}\widetilde{\alpha}\biggr)$ of
(\ref{sixtysix}) is exactly the one that can "pass through"\footnote{
By "{\it pass through}" we mean a procedure explained in appendix C.} 
the Hamiltonian $\HCT$ of KvN if we change the gauge field as we do in
(\ref{sevenb}). In that appendix we shall  also explore which gauge fields 
 have to be inserted in  $\HCT$ to allow for a 
general\footnote{By
"general" we mean one not of the form $\displaystyle exp\biggl(i\frac{e}{c}\widetilde{\alpha}\biggr)$.} phase
$\displaystyle exp[i\alpha(q,\lambda_p)]$ to "pass through"
$\HCT$.

\section{Landau Problem}

In this section we will make a first application of the minimal coupling scheme that we have developed 
previously for the KvN formalism. 
This first application is the Landau problem. We will first review it in quantum mechanics and then turn to
the classical operatorial version of KvN. The Landau problem is concerned with the dynamics of a particle under a 
constant magnetic field directed along
$z$. We make the following choice for the gauge potential:
%%%
\begin{equation}
A_x=0,\;\;\;\;\;A_y=Bx,\;\;\;\;A_z=0 \label{choice}
\end{equation}
%%%
The Schr\"odinger Hamiltonian is then
%%%
\begin{equation}
\displaystyle \widehat{H}=\frac{1}{2m}\biggl[\widehat{p}_x^{\,2}+\biggl(\widehat{p}_y-\frac{eB}{c}\widehat{x}\biggr)^2+
\widehat{p}_z^{\,2}\biggr]
\end{equation}
%%%
As $\widehat{p}_y,\widehat{p}_z$ commute with $\widehat{H}$ we can diagonalize all these three operators simultaneously. 
The eigenfunctions
will then be labelled by the eigenvalues of $\widehat{H}$, i.e. $E$, and by those of $\widehat{p}_y,\widehat{p}_z$ 
which are
$p^0_y$ and $p^0_z$. Their form will be
%%%
\begin{equation}
\displaystyle
\psi_{\s E,p^0_y,p^0_z}(x,y,z)=\frac{1}{2\pi\hbar}exp\biggl[\frac{i}{\hbar}(p_y^0y+p_z^0z)\biggr]\psi(x) \label{lannine}
\end{equation}
%%%
The stationary eigenvalue problem 
%%%
\begin{equation}
\widehat{H}\psi_{\s E,p^0_y,p^0_z}=E\psi_{\s E,p^0_y,p^0_z}
\end{equation}
%%%
leads to the following differential equation for $\psi(x)$
%%%
\begin{equation}
\displaystyle
-\frac{\hbar^2}{2m}\psi^{\prime\prime}(x)+\frac{1}{2m}\biggl(p_y^0-\frac{eBx}{c}\biggr)^2\psi(x)=\biggl(E-\frac{p^{0^2}_z}
{2m}\biggr)\psi(x) \label{diffeq}
\end{equation}
%%%
Indicating with $E_t$ the quantity
%%%
\begin{equation}
E_t\equiv E-\frac{p_z^{0^2}}{2m} \label{moden}
\end{equation}
%%%
and making a change of variables from $x$ to $x^{\prime}$, with $\displaystyle x^{\prime}\equiv p_y^0-\frac{eBx}{c}$,
eq. (\ref{diffeq}) is turned into the following one
%%%
\begin{equation}
\displaystyle
-\frac{\hbar^2}{2m}\psi^{\prime\prime}(x^{\prime})+\frac{1}{2m}\biggl[\frac{c}{eB}\biggr]^2x^{\prime^2}\psi(x^{\prime})
=\biggl[\frac{c}{eB}\biggr]^2E_t\psi(x^{\prime})
\end{equation}
%%%
We can immediately notice that this  is like an harmonic oscillator eigenvalue problem with the frequency replaced by
%%%
\begin{equation}
\omega\equiv\frac{c}{eBm} \label{modfreq}
\end{equation}
%%%
and with the energy replaced by $\displaystyle
\biggl[\frac{c}{eB}\biggr]^2E_t$.
This quantity is discretized like in the harmonic oscillator problem:
%%%
\begin{equation}
\biggl[\frac{c}{eB}\biggr]^2E_t=\hbar\omega\biggl(n+\frac{1}{2}\biggr) \label{modeigen}
\end{equation}
%%%
Combining (\ref{moden}) and (\ref{modfreq}) we get from (\ref{modeigen})
%%%
\begin{equation}
E_{n,p_z^0}=\frac{e\hbar B}{mc}\biggl(n+\frac{1}{2}\biggr)+\frac{p^{0^2}_z}{2m}
\end{equation}
%%%
So the eigenfunctions (\ref{lannine}) can be labelled by the quantum numbers $(n,p_y^0,p_z^0)$: i.e. $\psi_{n,p_y^0,p_z^0}$.
Note that these wave functions are degenerate because all those with different values of $p_y^0$ have the same 
$E_{n,p_z^0}$. So there is an infinite degeneracy.

Let us now analyze the same problem at the classical level using the operatorial formalism of KvN.
Using (\ref{abstham}) and the gauge choice (\ref{choice}), the $\HT$ is
%%%
\begin{equation}
\HT=\frac{1}{m}\lambda_xp_x+\frac{1}{m}\biggl(\lambda_y-\frac{e}{c}{\mathcal A}_y\biggr)\biggl(p_y-\frac{e}{c}A_y
\biggr)+\frac{1}{m}\lambda_zp_z
\end{equation}
%%%
We will now turn $\HT$ into an operator $\HCT$ using the "mixed" representation (\ref{mixrep}) and what 
we get is: 
%%%
\begin{equation}
\displaystyle
\HCT=\frac{1}{m}\frac{\partial}{\partial x}\frac{\partial}{\partial\lambda_{p_x}}+
\frac{1}{m}\biggl(-i\frac{\partial}{\partial y}+\frac{eB}{c}\lambda_{p_x}\biggr)
\biggl(i\frac{\partial}{\partial\lambda_{p_y}}-\frac{eB}{c}x\biggr)+
\frac{1}{m}\frac{\partial}{\partial z}\frac{\partial}{\partial\lambda_{p_z}} \label{mixham}
\end{equation}
%%%
Let us now diagonalize this operator. The reason to do that is because in the KvN theory the equation to solve is 
(\ref{lio2}) which, because of  (\ref{operat}),  can be written as
%%%
\begin{equation}
i\partial_t\psi=\hat{\HT}\psi
\end{equation}
%%%
So, like for the Schr\"odinger equation, one should first diagonalize $\HCT$
%%%
\begin{equation}
\HCT\,\psi_{\s \widetilde{E}}=\widetilde{E}\,\psi_{\s \widetilde{E}} \label{mixeigen}
\end{equation}
%%%
and then write a generic wave function as
%%%
\begin{equation}
\displaystyle
\psi(t)=\sum_{\s \widetilde{E}}C_{\s \widetilde{E}}e^{-i\widetilde{E}t}\psi_{\s \widetilde{E}} \label{sevensix}
\end{equation}
%%%
where the $C_{\widetilde{E}}$ are derived from the expansion of the initial $\psi$ on the $\psi_{\s \widetilde{E}}$.
We want to underline that the $\widetilde{E}$ that appear in eqs. (\ref{mixeigen})-(\ref{sevensix}) {\it have nothing
to do with the physical energy} of the system. They are simply the possible eigenvalues of the evolution operator
$\HCT$ and, using them and the associated eigenfunctions, we can deduce the evolution of the $\psi$ like it is done 
in formula (\ref{sevensix}).
 For more details about this and on the 
manner to reconstruct the standard deterministic motion of CM see ref. \cite{Mauro}.

Let us now turn to (\ref{mixham}) and diagonalize it like in (\ref{mixeigen}). Note that the operators:
$\displaystyle -i\frac{\partial}
{\partial y}, i\frac{\partial}{\partial\lambda_{p_y}},-i\frac{\partial}{\partial z}$ and $\displaystyle
i\frac{\partial}{\partial
\lambda_{p_z}}$ commute with
$\HCT$ and so we can diagonalize  simultaneously these five operators. The generic eigenfunction (in the mixed
representation
$q,\lambda_p$) has the form 
%%%
\begin{equation}
\displaystyle
\psi(q,\lambda_p)=\frac{1}{(2\pi)^2}exp[i\lambda_y^0y-i\lambda_{p_y}p_y^0]exp[i\lambda_z^0z-i\lambda_{p_z}p_z^0]
\psi(x,\lambda_{p_x}) \label{sevsev}
\end{equation}
%%%
where $\lambda_y^0, p_y^0, \lambda_z^0$ and $p_z^0$ are eigenvalues of the operators $\displaystyle
-i\frac{\partial}{\partial y},
i\frac{\partial}{\partial\lambda_{p_y}}, -i\frac{\partial}{\partial z}, i\frac{\partial}{\partial\lambda_{p_z}}$
respectively. We see the similarity with the quantum case except for the fact that the dimension of the space is double.
Inserting (\ref{sevsev}) in (\ref{mixeigen}) we get the equation
%%%
\begin{equation}
\displaystyle
\biggl[\frac{1}{m}\frac{\partial}{\partial x}\frac{\partial}{\lambda_{p_x}}+\frac{1}{m}\biggl(\lambda_y^0+\frac{eB}{c}
\lambda_{p_x}\biggr)\biggl(p_y^0-\frac{eB}{c}x\biggr)+\frac{1}{m}\lambda_z^0p_z^0\biggr]\psi(x,\lambda_{p_x})
=\widetilde{E}\,\psi(x,\lambda_{p_x}) \label{lanotto}
\end{equation}
%%%
Via the new quantity
%%%
\begin{equation}
\widetilde{E}_+\equiv\widetilde{E}-\frac{1}{m}\lambda_z^0p_z^0 \label{enplus}
\end{equation}
%%%
we can rewrite eq. (\ref{lanotto}) as
%%%
\begin{equation}
\displaystyle
\biggl[\frac{1}{m}\frac{\partial}{\partial x}\frac{\partial}{\partial\lambda_{p_x}}+\frac{1}{m}\biggl(\lambda_y^0+
\frac{eB}{c}\lambda_{p_x}\biggr)\biggl(p_y^0-\frac{eB}{c}x\biggr)\biggl]\psi(x,\lambda_{p_x})
=\widetilde{E}_+\,\psi(x,\lambda_{p_x}) \label{lannove}
\end{equation}
%%%
Doing now the following change of variables
%%%
\begin{eqnarray}
&&x^{\prime}\equiv x-\frac{c}{eB}p_y^0\nonumber\\
&&\lambda_{p_x}^{\prime}\equiv\lambda_{p_x}+\frac{c}{eB}\lambda_y^0
\end{eqnarray}
%%%
we can rewrite eq. (\ref{lannove}) as
%%%
\begin{equation}
\biggl[\frac{1}{m}\frac{\partial}{\partial x^{\prime}}\frac{\partial}{\partial\lambda_{p_x}^{\prime}}
-\frac{1}{m}\biggl(\frac{eB}{c}\biggr)^2\lambda_{p_x}^{\prime}x^{\prime}\biggr]\psi(x^{\prime},\lambda_{p_x}^{\prime})
=\widetilde{E}_+\,\psi(x^{\prime},\lambda_{p_x}^{\prime})
\end{equation}
%%%
The dimensions of the various quantities are such that we can write the above equation as 
%%%
\begin{equation}
\biggl[\frac{1}{m}\frac{\partial}{\partial x^{\prime}}\frac{\partial}{\partial\lambda_{p_x}^{\prime}}
-m\omega^2\lambda_{p_x}^{\prime}x^{\prime}\biggr]\psi(x^{\prime},\lambda_{p_x}^{\prime})
=\widetilde{E}_+\,\psi(x^{\prime},\lambda_{p_x}^{\prime}) \label{lannovebis}
\end{equation}
%%%
where $\displaystyle \omega\equiv\frac{eB}{mc}$ has the dimension of an angular velocity and it is related to the
well-known  Larmor frequency of rotation of a particle in a constant magnetic field.
Eq. (\ref{lannovebis}) is the KvN eigenvalue equation for an harmonic oscillator which is studied in details in 
appendix D which we advice the reader to go through before going on. 
The spectrum of the $\HCT$ for the harmonic oscillator (see appendix D)  is given by\footnote{Note that we have
a discretization phenomenon even at the {\it classical} level for the eigenvalues of the Liouvillian. This is related to
the requirement of single valuedness of the KvN states as explained in appendix D.}
%%%
\begin{equation}
\widetilde{E}_{\s +N}^{osc}=N\omega,\;\;\;\;\;\;N=\cdots,-2,-1,0,1,2,\cdots
\end{equation}
%%%
So, using (\ref{enplus}), the final spectrum of eq. (\ref{lanotto}) is  
%%%
\begin{equation}
\widetilde{E}=N\biggl(\frac{eB}{mc}\biggr)+\frac{1}{m}\lambda_z^0p_z^0
\end{equation}
%%%
and the wave functions are any linear combination of the following eigenfunctions (see appendix D):
%%%
\begin{equation}
\displaystyle
\psi_{N,n,\lambda_y^0,p_y^0,\lambda_z^0,p_z^0}=\frac{1}{2\pi}exp[i\lambda_y^0y-i\lambda_{p_y}p_y^0]
exp[i\lambda_z^0z-i\lambda_{p_z}p_z^0]\cdot\psi_n^{osc}(Z_+)\psi_{n+N}^{osc}(Z_-) \label{liouvham}
\end{equation}
%%%
where $\psi_n^{osc}$ are the eigenfunctions of the {\it quantum} 1-dim harmonic 
oscillator\footnote{$n=-N, -N+1, -N+2,\cdots$ if $N$ is negative and $n=0, 1, 2,\cdots$
if $N$ is positive.}
with $\hbar$
replaced by an  arbitrary quantity $\Delta$ which has the dimension of an action and with $Z_+,Z_-$ defined as :
%%%
\begin{equation}
Z^+=\frac{x^{\prime}+\Delta\lambda_{p_x}^{\prime}}{\sqrt{2}},\;\;\;\;
Z^-=\frac{x^{\prime}-\Delta\lambda_{p_x}^{\prime}}{\sqrt{2}}
\end{equation}
%%%
We see from the form of the wave functions in (\ref{liouvham}) 
that the degeneracy is much more than in the quantum case. Not only 
the eigenfunctions with different 
values of $p_y^0$ have the same  $\widetilde{E}$, but the same happens
for those  eigenfunctions with different values\footnote{with the only constraint that the product $\lambda_z^0p_z^0$ be the same.}
of $\lambda_y^0$, $n$
and of 
$\lambda_z^0, p_z^0$.
So it is a much more wider degeneracy
than in the quantum case. The reader may wonder why this happens.
We feel that this may be due to the fact that the "wave functions" in the KvN formalism have a
number of variables $(q,\lambda_p)$ that is double than in QM.

The reader may be puzzled by the presence in the eigenfunctions (\ref{liouvham}) of the 
arbitrary quantity $\Delta$ which did not appear in the original $\HT$. This is not a problem. In fact once we are given 
an initial wave function $\widetilde{\psi}(q,\lambda_p)$ not depending on $\Delta$, we will expand it on the basis
(\ref{liouvham}) as:
%%%
\begin{equation}
\widetilde{\psi}(q,\lambda_p)=\sum_{N,n,\cdots}\biggl(C_{N,n,\lambda_y^0,p_y^0,\lambda_z^0,p_z^0}\biggr)
\biggl(\psi_{N,n,\lambda_y^0,p_y^0,\lambda_z^0,p_z^0}\biggr)
\end{equation}
%%%
where $C_{N,n\cdots}$ are  coefficients which will depend on $\Delta$ and this dependence  will compensate the one contained 
in the $\psi_{N,n\cdots}$. The evolution in $t$ will not reintroduce the dependence on $\Delta$ because the eigenvalues
of $\HCT$ do not depend on $\Delta$ as shown in appendix D. Solving a classical system via its KvN states is like
working in the Schr\"odinger picture. We could also work out the analog of the Heisenberg picture and this
is done for the Landau problem in appendix E.

\section{Aharonov-Bohm Phenomenon}

\begin{figure}
\centering
\includegraphics{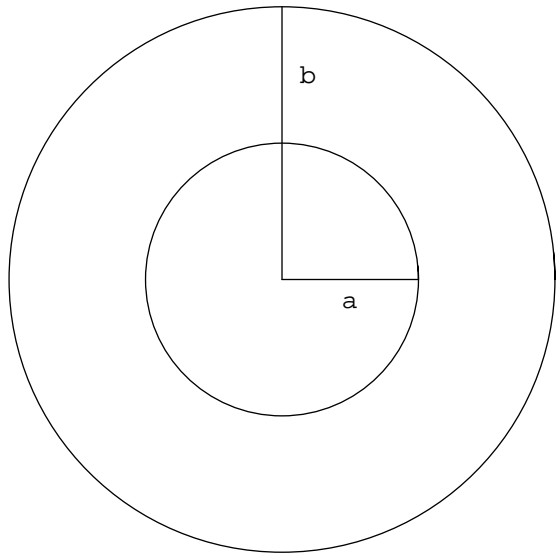}
\caption{Aharonov-Bohm geometrical set up} \label{ABset}
\end{figure}

The second application of the MC that we will study is the well-known Aharonov-Bohm (AB) effect, \cite{Bohm}. This is a
phenomenon which proves that the QM wave functions are changed by the presence of a gauge potential even in
regions where the magnetic 
field associated to this gauge potential is zero. The change in the wave functions can be detected by an interference 
experiment. The classical motion instead feels only the magnetic field and not the gauge potential. 
In this section we will study this 
phenomenon not by looking at wave functions but at the spectrum of respectively the Schr\"odinger operator $\widehat{H}$
and the classical KvN Liouville operator $\HCT$. The geometrical set up that we will use for the AB effect is
illustrated in  Figure  \ref{ABset} and it has been suggested in the book of Sakurai \cite{quantumbooks}. 
Basically we have two
infinitely long cylinders one inside the other. We will study the Schr\"odinger operator $\widehat{H}$ in the
region in between the two cylinders. We will show, as Sakurai \cite{quantumbooks} indicated, that {\it 
the spectrum of $\widehat{H}$ changes once we turn on the magnetic field} inside the smaller cylinder. In the region in 
between the two cylinders the magnetic field is zero because the smaller cylinder shields completely the magnetic field. So
we are exactly in an AB configuration: zero magnetic field and non-zero gauge potential. Using
the same geometrical configuration, we will study the spectrum of the KvN-Liouville operator $\HCT$ and we will
prove that {\it it does not change} once we turn on the magnetic field differently than what happens in QM.
We feel that this, in the framework of the operatorial formulation of CM, is the best mathematical proof that there is no
AB effect in CM.

\begin{figure}
\centering
\includegraphics{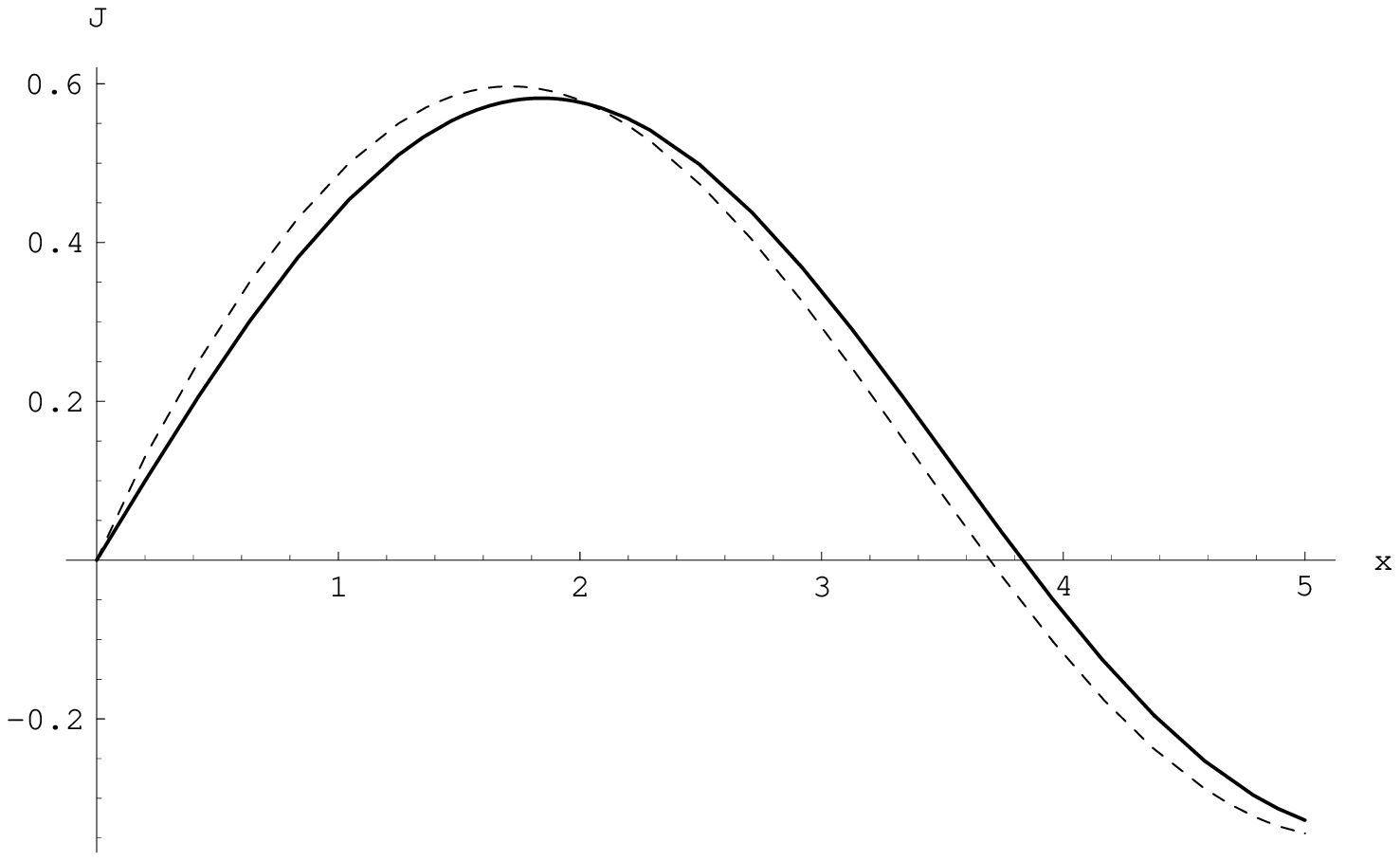}
\caption{Zeros of Bessel functions: m=1 (continuous line), m=0.9 (dashed line)} \label{Bes}
\end{figure}

Let us now study the Schr\"odinger operator in the geometrical set up of Figure \ref{ABset} and let us do it at first 
without magnetic field. The Schr\"odinger operator in cylindrical coordinates
%%%
\begin{equation}
\label{cylindrical}
\left\{
	\begin{array}{l}
	\displaystyle x=\rho \,cos\varphi\smallskip\\
          \displaystyle y=\rho \,sin\varphi\smallskip\\
          \displaystyle z=z\\
	\end{array}
	\right.
\end{equation}
%%%
is for a free particle:
%%%
\begin{equation}
\displaystyle
\widehat{H}=-\frac{\hbar^2}{2\mu}\biggl(\frac{\partial^2}{\partial\rho^2}+\frac{1}{\rho}\frac{\partial}{\partial\rho}
+\frac{1}{\rho^2}\frac{\partial^2}{\partial\varphi^2}+\frac{\partial^2}{\partial z^2}\biggr) \label{freeH}
\end{equation}
%%%
and the eigenvalues equation is
%%%
\begin{equation}
\displaystyle
-\frac{\hbar^2}{2\mu}\biggl(\frac{\partial^2}{\partial\rho^2}+\frac{1}{\rho}\frac{\partial}{\partial\rho}
+\frac{1}{\rho^2}\frac{\partial^2}{\partial\varphi^2}+\frac{\partial^2}{\partial z^2}\biggr)\psi(\rho,\varphi,z)
=E\,\psi(\rho,\varphi,z) \label{bessel1}
\end{equation}
%%%
As the operators $\displaystyle 
\frac{\partial}{\partial\varphi}$ and $\displaystyle \frac{\partial}{\partial z}$ commute with $\widehat{H}$ we can diagonalize
these three operators simultaneously and have
%%%
\begin{equation}
\displaystyle
\psi(\rho,\varphi,z)=\frac{1}{2\pi}exp\biggl[\frac{ip_z^0z}{\hbar}\biggr]exp[im\varphi]R(\rho) \label{decomp1}
\end{equation}
%%%
where $p_z^0$ is a fixed value and $m$ is an integer. Inserting (\ref{decomp1}) in (\ref{bessel1}) we get the following
equation for $R(\rho)$
%%%
\begin{equation}
R^{\prime\prime}(\rho)+\frac{R^{\prime}(\rho)}{\rho}+\biggl(\bar{s}-\frac{m^2}{\rho^2}\biggr)R(\rho)=0 \label{besselr}
\end{equation}
%%%
where 
%%%
\begin{equation}
\bar{s}\equiv\frac{2\mu E}{\hbar^2}-\frac{p_z^{0^2}}{\hbar^2} \label{abar}
\end{equation}
%%%
If $\bar{s}=1$ then eq. (\ref{besselr}) would be the well-known Bessel equation \cite{Watson}. 
We can get it by using the new variables 
$r\equiv\sqrt{\bar{s}}\rho$. In "$r$" eq. (\ref{besselr}) is
%%%
\begin{equation}
\displaystyle
\frac{\partial^2R}{\partial r^2}+\frac{1}{r}\frac{\partial R}{\partial r}+\biggl(1-\frac{m^2}{r^2}\biggr)R=0
\label{besselr2}
\end{equation}
%%%
As we want the wave function to be confined 
between the two cylinders, the boundary conditions should be
%%%
\begin{equation}
R(\sqrt{s}a)=R(\sqrt{s}b)=0 \label{bouncon}
\end{equation}
%%%
where $a$ and $b$ are the radii of respectively the smaller and the larger
cylinder, see Figure \ref{ABset}. 
The general solution of eq. (\ref{besselr2}), {\it with m integer}, is given by the linear 
combination
of the Bessel functions of the first and second kind which are \cite{Watson}:
%%%
\begin{eqnarray}
\displaystyle
&&J_m(r)=\biggl(\frac{r}{2}\biggr)^m\sum_{n=0}^{\infty}\frac{(-1)^n\bigl(\frac{r}{2}\bigr)^{2n}}{n!\,\Gamma(n+m+1)}
\nonumber\\
&&Y_m(r)=\lim_{\epsilon\to 0}\frac{1}{\epsilon}[J_{m+\epsilon}(r)-(-1)^mJ_{-m-\epsilon}(r)]
\end{eqnarray}
%%%
The general solution will then be
%%%
\begin{equation}
R(\sqrt{\bar{s}}\rho)=AJ_m(\sqrt{\bar{s}}\rho)+BY_m(\sqrt{\bar{s}}\rho) \label{solutions}
\end{equation}
%%%
Imposing the boundary conditions (\ref{bouncon}) we will get the spectrum of the system. In order to simplify things 
we will
consider the limiting case in which the radius of the internal cylinder $a$ goes to zero. In this case the boundary 
conditions
(\ref{bouncon}) become
%%%
\begin{equation}
R(0)=0, \;\;\;\; R(\sqrt{\bar{s}}b)=0 \label{bouncon2}
\end{equation}
%%%
It is well known \cite{Watson} that the Bessel functions of the second  kind $Y$ are singular in the origin, so we 
restrict ourselves
to solutions (\ref{solutions}) of the form:
%%%
\begin{equation}
R(\sqrt{\bar{s}}\rho)=AJ_m(\sqrt{\bar{s}}\rho)
\end{equation}
%%%
With this choice the first of the boundary conditions (\ref{bouncon2}) is automatically satisfied because 
\cite{Watson} $J_m(0)=0$
for $m\ge 1$. We have to satisfy only the second one of the conditions (\ref{bouncon2}) which implies:
%%%
\begin{equation}
J_m(\sqrt{\bar{s}}b)=0 \label{last}
\end{equation}
%%%
This relation tells us that we have to look for the zeros of the above Bessel functions. Let us call them 
$\alpha_{k,m}$ where $m$ indicates to which 
Bessel
function we refer to and $k$ labels the various zeros of the $m$-Bessel function in increasing order, so $k=1,2,3,\cdots$.
The solutions of eq. (\ref{last}) can then be formally written as 
%%%
\begin{equation}
\sqrt{\bar{s}}b=\alpha_{k,m}
\end{equation}
%%%
Replacing $\bar{s}$ in the equation above with its expression (\ref{abar}), we get that
%%%
\begin{equation}
\displaystyle
E_{k,m}=\hbar^2\frac{\alpha^2_{k,m}}{2\mu b^2}+\frac{p_z^{0^2}}{2\mu}
\end{equation}
%%%
These are the energy levels. If we choose $m=1$ and the second zero ($k=2$)  
which is $\alpha_{2,1}
\approx 3.83$, we get
%%%
\begin{equation}
E_{2,1}=\hbar^2\frac{7.33}{\mu b^2}+\frac{p_z^{0^2}}{2\mu} \label{edueuno}
\end{equation}
see Figure \ref{Bes}.

Let us now turn on the magnetic field \cite{Bohm} which we want to be zero everywhere except for the $B_z$ component on
the line $x^2+y^2=0$ and with a fixed flux $\Phi_{\s B}$. A choice of the gauge potential \cite{Bohm} is
%%%
\begin{equation}
\left\{
	\begin{array}{l}
	\displaystyle A_x=\frac{-y\Phi_{\s B}}{2\pi(x^2+y^2)}\smallskip\\
          \displaystyle A_y=\frac{x\Phi_{\s B}}{2\pi(x^2+y^2)}\smallskip\\
          \displaystyle A_z=0 \label{Abohm}\\
	\end{array}
	\right.
\end{equation}
%%%
Turning to cylindrical coordinates we have that (\ref{Abohm}) is equivalent to 
%%%
\begin{equation}
A_{\rho}=0,\;\;\;A_{\varphi}=\frac{\Phi_{\s B}}{2\pi\rho},\;\;\; A_z=0
\end{equation}
%%%
So with the above choice of gauge the minimal coupling affects only the 
$\displaystyle -i\hbar\frac{\partial}{\partial\varphi}$:
%%%
\begin{equation}
-i\hbar\frac{\partial}{\partial\varphi}\;\longrightarrow\;-i\hbar\frac{\partial}{\partial\varphi}-\frac{e}{c}\frac{\Phi_
{\s B}}{2\pi}
\end{equation}
%%%
The associated Schr\"odinger operator (\ref{freeH}) is then: 
%%%
\begin{equation}
\displaystyle
\widehat{H}\;\longrightarrow\;\widehat{H}_{\s B}=\frac{-\hbar^2}{2\mu}\biggl[\frac{\partial^2}{\partial\rho^2}
+\frac{1}{\rho}\frac{\partial}{\partial\rho}+\frac{1}{\rho^2}\biggl(\frac{\partial}{\partial\varphi}-\frac{ie}{ch}
\Phi_{\s B}\biggr)^2+\frac{\partial^2}{\partial z^2}\biggr]
\end{equation}
%%%
and the eigenvalue equation is
%%%
\begin{equation}
\displaystyle
\frac{-\hbar^2}{2\mu}\biggl[\frac{\partial^2}{\partial\rho^2}
+\frac{1}{\rho}\frac{\partial}{\partial\rho}+\frac{1}{\rho^2}\frac{\partial^2}{\partial\varphi^2}
-\frac{2ie}{ch}\Phi_{\s B}\frac{1}{\rho^2}\frac{\partial}{\partial\varphi}-\frac{e^2}{c^2h^2}
\frac{\Phi^2_{\s B}}{\rho^2}+\frac{\partial^2}{\partial z^2}\biggr]\psi=E\psi(\rho,\varphi,z) \label{ninenine}
\end{equation}
%%%
Like  for  (\ref{decomp1}) we can choose solutions of the form
%%%
\begin{equation}
\displaystyle
\psi(\rho,\varphi,z)=\frac{1}{2\pi}exp\biggl[\frac{ip_z^0z}{\hbar}\biggr]exp[im\varphi]R(\rho)
\end{equation}
%%%
which, inserted in (\ref{ninenine}), give the following differential equation for $R(\rho)$
%%%
\begin{equation}
\displaystyle
-\frac{\hbar^2}{2\mu}\biggl[\frac{R^{\prime\prime}}{R}+\frac{1}{\rho}\frac{R^{\prime}}{R}-\frac{1}{\rho^2}
\biggl(m-\frac{e\Phi_{\s B}}{ch}\biggr)^2-\frac{p_z^{0^2}}{\hbar^2}\biggr]=E \label{eigeneq}
\end{equation}
%%%
Indicating with $\displaystyle \alpha\equiv\frac{e\Phi_{\s B}}{ch}$ and with
$\displaystyle \bar{s}=\frac{2\mu E}{\hbar^2}-\frac{p_z^{0^2}}{\hbar^2}$, eq. (\ref{eigeneq}) can be written as 
%%%
\begin{equation}
R^{\prime\prime}(\rho)+\frac{R^{\prime}(\rho)}{\rho}+\biggl[\bar{s}-\frac{(m-\alpha)^2}{\rho^2}\biggr]R(\rho)=0
\label{besselmag}
\end{equation}
%%%
If we compare the previous equation with (\ref{besselr}) we notice that having turned on the magnetic field has only shifted 
$m\rightarrow m_{\s B}\equiv m-\alpha$. Here $m_{\s B}$ will not be anymore an integer but a real number.
Doing the same change of variables as before, $r\equiv\sqrt{\bar{s}}\rho$, we can transform (\ref{besselmag}) into
%%%
\begin{equation}
\frac{\partial^2 R}{\partial r^2}+\frac{1}{r}\frac{\partial R}{\partial r}+\biggl(1-\frac{m^2_{\s B}}{r^2}\biggr)R=0
\label{besselmag2}
\end{equation}
%%%
For this equation with $m_{\s B}$ real there are two linearly independent solutions which are two Bessel functions of the 
first kind with opposite indeces:
%%%
\begin{eqnarray}
\displaystyle
&& J_{m_{\s B}}(r)=\biggl(\frac{r}{2}\biggr)^{m_{\s B}}\sum_{n=0}^{\infty}\frac{(-1)^n\bigl(\frac{r}{2}\bigr)^{2n}}
{n!\,\Gamma(n+m_{\s B}+1)}\nonumber\\
&& J_{-m_{\s B}}(r)=\biggl(\frac{r}{2}\biggr)^{-m_{\s B}}\sum_{n=0}^{\infty}\frac{(-1)^n\bigl(\frac{r}{2}\bigr)^{2n}}
{n!\,\Gamma(-n+m_{\s B}+1)}
\end{eqnarray}
%%%
One immediately notices that $J_{m_{\s B}}(0)=0$ while $J_{-{m_{\s B}}}(0)$ diverges.
As before we must have as boundary conditions $J_{m_{\s B}}(0)=0$ and so the general solution of (\ref{besselmag2}) is
%%%
\begin{equation}
R(r)=AJ_{m_{\s B}}(r)
\end{equation}
%%%
The other boundary condition gives the following relation
%%%
\begin{equation}
J_{m_{\s B}}(\sqrt{\bar{s}}b)=0
\end{equation}
%%%
from which, as before, we can derive the energy levels
%%%
\begin{equation}
\displaystyle
E_{k,m_{\s B}}=\hbar\frac{\alpha^2_{k,m_{\s B}}}{2\mu b^2}+\frac{p_z^{0^2}}{2\mu} \label{ekappa}
\end{equation}
%%%
If $\Phi_{\s B}$ is such to give, for example, $\alpha=0.1$, then we will have to consider 
the Bessel functions $J_{m-0.1}$. The second zero of $J_{0.9}$, analog
to the second one of $J_1$ that we considered before, is $\alpha_{2, 0.9}=3.70$, 
see Figure \ref{Bes}.
Inserting this value  in (\ref{ekappa}) we get
%%%
\begin{equation}
\displaystyle
E_{2, 0.9}=\hbar^2\frac{6.84}{\mu b^2}+\frac{p_z^{0^2}}{2\mu}<E_{2,1}
\end{equation}
%%%
The last inequality indicates that $E_{2,0.9}$ is smaller than the corresponding level $E_{2,1}$ of the case without magnetic field
calculated in (\ref{edueuno}). So this is a clear indication, like it was suggested in ref. \cite{quantumbooks}, that the presence
of a gauge potential modifies the spectrum of the Schr\"odinger operator even if the wave function is restricted to an
area with zero magnetic field. 

Let us now perform the same analysis in the classical case using the KvN operator $\HCT$. We have to be careful here because,
we have to go to cylindrical coordinates both for the $q$ and the $p$ variables and for their derivatives
$\displaystyle \frac{\partial}{\partial q},\frac{\partial}{\partial p}$ which enter $\HCT$. In the case without magnetic
field the Lagrangian is 
%%%
\begin{equation}
L=\frac{1}{2}\mu\dot{\rho}^2+\frac{1}{2}\mu\rho^2\dot{\varphi}^2+\frac{1}{2}\mu\dot{z}^2
\end{equation}
%%%
and so the momenta conjugate to $\rho,\varphi,z$ are:
%%%
\begin{equation}
\left\{
	\begin{array}{l}
	\displaystyle p_{\rho}=\mu\dot{\rho}\smallskip\\
          \displaystyle p_{\varphi}=\mu\rho^2\dot{\varphi}\smallskip\\
          \displaystyle p_z=\mu\dot{z}\\
	\end{array}
	\right.
\end{equation}
%%%
The relations between $p_x,p_y,p_z$ and $p_{\rho},p_{\varphi},p_z$  can be easily worked: 
%%%
\begin{equation}
\left\{
	\begin{array}{l}
	\displaystyle p_x=\mu\dot{x}=\mu\dot{\rho}cos\varphi-\mu\rho\dot{\varphi}sin\varphi=
	p_{\rho}cos\varphi-\frac{p_{\varphi}}{\rho}
	sin\varphi\smallskip\\
          \displaystyle p_{y}=\mu\dot{y}=\mu\dot{\rho}sin\varphi+\mu\rho\dot{\varphi}cos\varphi=p_{\rho}sin\varphi+\frac{p_{\varphi}}{\rho}
	cos\varphi\smallskip\\
          \displaystyle p_z=p_z\\
	\end{array}
	\right.
\end{equation}
%%%
As the $z$ and $p_z$ are the same in the two coordinate systems we can summarize the basic transformations in the
following set
%%%
\begin{equation}
\left\{
	\begin{array}{l}
	\displaystyle x=\rho cos\varphi\smallskip\\
          \displaystyle y=\rho sin\varphi\smallskip\\
          \displaystyle p_x=p_{\rho}cos\varphi-\frac{p_{\varphi}}{\rho}sin\varphi\smallskip\\
          \displaystyle p_y=p_{\rho}sin\varphi+\frac{p_{\varphi}}{\rho}cos\varphi \label{cyl}\\
	\end{array}
	\right.
\end{equation}
%%%
We should note that the transformations of the momenta $p_x,p_y$ are not just functions of the new momenta $p_{\rho},p_{
\varphi}$ but also of the cylindrical coordinates $\varphi,\rho$. Let us remember that $\HCT$ 
in cartesian coordinates contains the derivatives
$\displaystyle \frac{\partial}{\partial q},\frac{\partial}{\partial p}$ and so we should check how they 
are related to the derivatives in cylindrical coordinates. Using (\ref{cyl}) it is a long but easy calculation to show that:
%%%
\begin{equation}
\displaystyle
\pmatrix{\partial/\partial x\cr
               \partial/\partial y\cr
               \partial/\partial p_x\cr
               \partial/\partial p_y\cr}=
                    \pmatrix{cos\varphi & -sin\varphi/\rho & -p_{\varphi}sin\varphi/\rho^2 &
                    p_{\varphi}/\rho\cdot cos\varphi+p_{\rho}sin\varphi\cr
                    sin\varphi & cos\varphi/\rho & p_{\varphi}cos\varphi/\rho^2 &
                    p_{\varphi}/\rho\cdot sin\varphi-p_{\rho}cos\varphi\cr
                    0 & 0 & cos\varphi & -\rho sin\varphi\cr
                    0 & 0 & sin\varphi & \rho cos\varphi\cr}
                    \cdot\pmatrix{\partial/\partial \rho\cr
                    \partial/\partial \varphi \cr
                    \partial/\partial p_{\rho}\cr
                    \partial/\partial p_{\varphi}\cr} \label{cyl2}
                    \end{equation}
%%%
Equipped with these transformations we can then easily transform $\HCT$ from cartesian coordinates to cylindrical ones. 
In the  case of a free particle we get
%%%
\begin{eqnarray}
&&\HCT=-i\frac{p_x}{\mu}\frac{\partial}{\partial x}-i\frac{p_y}{\mu}\frac{\partial}{\partial y}
-i\frac{p_z}{\mu}\frac{\partial}{\partial z}=\nonumber\\
&&=-\frac{i}{\mu}p_{\rho}\frac{\partial}{\partial\rho}-\frac{i}{\mu}\frac{p_{\varphi}}{\rho^2}\frac{\partial}{\partial\varphi}
-\frac{i}{\mu}p_z\frac{\partial}{\partial z}-\frac{i}{\mu}\frac{p_{\varphi}^2}{\rho^3}\frac{\partial}{\partial p_{\rho}}
\label{freecyl}
\end{eqnarray}
%%%
where in the second step we have used (\ref{cyl}) and (\ref{cyl2}). Next we will turn on the magnetic field whose
gauge potential is in (\ref{Abohm}). The minimal coupling for $\HT$ is given in (\ref{abstham}) and we need it to 
build the ${\mathcal A}_{q_i}$. For our potential these ${\mathcal A}_{q_i}$ are 
%%%
\begin{equation}
\label{gouno}
\left\{
	\begin{array}{l}
	\displaystyle {\mathcal A}_x=-\lambda_{p_x}\frac{\Phi_{\s B}}{\pi}\frac{xy}{(x^2+y^2)^2}+\lambda_{p_y}\frac{\Phi_{\s B}}{2\pi}
	\frac{x^2-y^2}{(x^2+y^2)^2}\smallskip\\
          \displaystyle 
          {\mathcal A}_y=\lambda_{p_y}\frac{\Phi_{\s B}}{2\pi}\frac{x^2-y^2}{(x^2+y^2)^2}+\lambda_{p_y}\frac{\Phi_{\s B}}{\pi}
	\frac{xy}{(x^2+y^2)^2}\smallskip\\
          \displaystyle {\mathcal A}_z=0\\
	\end{array}
	\right.
\end{equation}
%%%
In the expression above we have now to turn the $\lambda_{p_x},\lambda_{p_y}$ into operators, like in (\ref{oprep}), 
and next we have to change everything into 
cylindrical coordinates using (\ref{cyl}) and (\ref{cyl2}). The result is
%%%
\begin{equation}
\label{godue}
\left\{
	\begin{array}{l}
	\displaystyle \widehat{{\mathcal A}}_x=i\frac{\Phi_{\s B}}{2\pi}\frac{1}{\rho^2}\biggl[sin\varphi\frac{\partial}{\partial p_{\rho}}
	-\rho cos\varphi\frac{\partial}{\partial p_{\varphi}}\biggr]\smallskip\\
          \displaystyle \widehat{{\mathcal A}}_y=-i\frac{\Phi_{\s
B}}{2\pi}\frac{1}{\rho^2}\biggl[cos\varphi\frac{\partial}{\partial p_{\rho}}
	+\rho sin\varphi\frac{\partial}{\partial p_{\varphi}}\biggr]\smallskip\\
          \displaystyle \widehat{{\mathcal A}}_z=0\\
	\end{array}
	\right.
\end{equation}
%%%
Inserting (\ref{gouno})-(\ref{godue}) into (\ref{abstham}) and turning the $\lambda_{q_i}$ into operators, after a long but trivial 
calculation, we get
%%%
\begin{equation}
\displaystyle
\HCT_{\s A}\equiv -\frac{i}{\mu}p_{\rho}\frac{\partial}{\partial \rho}-\frac{i}{\mu\rho^2}\biggl(p_{\varphi}
-\frac{e\Phi_{\s B}}{2\pi c}\biggr)\frac{\partial}{\partial\varphi}-\frac{i}{\mu}p_z\frac{\partial}{\partial
z}-\frac{i}{\mu\rho^3}\biggl(p_{\varphi}-
\frac{e\Phi_{\s B}}{2\pi c}\biggr)^2\frac{\partial}{\partial p_{\rho}}
\label{oneoneone}
\end{equation}
%%%
Notice that we could obtain this $\HCT_{\s A}$ from the $\HCT$ of eq. (\ref{freecyl}) by just doing the
replacement
%%%
\begin{equation}
p_{\varphi}\;\longrightarrow\;p_{\varphi}-\frac{e\Phi_{\s B}}{2\pi c} \label{finiu}
\end{equation}
%%%
The eigenvalue equation in the free case
%%%
\begin{equation}
\HCT\,\psi(\rho,p_{\rho},\varphi,p_{\varphi},z,p_z)=\widetilde{E}\,\psi(\rho,p_{\rho},\varphi,p_{\varphi},z,p_z)
\label{psio}
\end{equation}
%%%
can be solved by noticing that $\HCT$ commutes with the four operators
$\displaystyle -i\frac{\partial}{\partial\varphi}$,$\displaystyle -i\frac{\partial}
{\partial z},p_{\varphi},p_z$. So these five operators can be 
diagonalized simultaneously and the solution will have the 
form
%%%
\begin{equation}
\displaystyle
\psi=\frac{1}{2\pi}\widetilde{R}(\rho,p_{\rho})\delta(p_{\varphi}-p_{\varphi}^0)\delta(p_z-p_z^0)exp(in\varphi)exp(i\lambda_z^0 z)
\label{psir}
\end{equation}
%%%
where $n$ is an integer and $p^0_{\varphi},p_z^0,\lambda_z^0$ are the eigenvalues of $\displaystyle \widehat{p}_{\varphi},
\widehat{p}_z,-i\frac{\partial}{\partial z}$. Inserting (\ref{psir}) in (\ref{psio}) we get the following 
equation for $\widetilde{R}
(\rho,p_{\rho})$
%%%
\begin{equation}
\biggl(-\frac{i}{\mu}p_{\rho}\frac{\partial}{\partial\rho}+\frac{p_{\varphi}^0n}{\mu\rho^2}-\frac{i}{\mu\rho^3}
p^{0^2}_{\varphi}\frac{\partial}{\partial
p_{\rho}}+\frac{\lambda_z^0p_z^0}{\mu}-\widetilde{E}\biggr)\widetilde{R}(\rho,p_{\rho})=0
\label{besselfree}
\end{equation}
%%%
We showed before that we could turn $\HCT$ into the $\HCT_{\s A}$ of (\ref{oneoneone}) by just doing the 
substitution (\ref{finiu}) $\displaystyle p_{\varphi}\rightarrow p_{\varphi}-\frac{e\Phi_{\s B}}{2\pi c}$. 
It is then clear
that we can turn also the solutions (\ref{psir}) into the solutions of the eigenvalue equation
%%%
\begin{equation}
\HCT_{\s A}\psi_{\s A}=\widetilde{E}_{\s A}\psi_{\s A} \label{sedici}
\end{equation}
%%%
by just doing the substitution (\ref{finiu}) into (\ref{psir}). The result is:
%%%
\begin{equation}
\displaystyle
\psi_{\s A}=\frac{1}{2\pi}\widetilde{R}_{\s A}(\rho,p_{\rho})\delta\biggl(p_{\varphi}-p_{\varphi}^0-\frac{e\Phi_{\s B}}{2\pi
c}\biggr)
\delta(p_z-p_z^0) exp(in\varphi)exp(i\lambda_z^0z) \label{psir2}
\end{equation}
%%%
Once we insert this into eq. (\ref{sedici}) we will get for $\widetilde{R}_{\s A}(\rho,p_{\rho})$ the following
equation:
%%%
\begin{equation}
\displaystyle
\biggl(-\frac{i}{\mu}p_{\rho}\frac{\partial}{\partial \rho}+\frac{p^{0}_{\varphi}n}{\mu\rho^2}-\frac{i}{\mu\rho^3}
p_{\varphi}^{0^2}\frac{\partial}{\partial p_{\rho}}+\frac{\lambda_z^0p_z^0}{\mu}-\widetilde{E}_{\s
A}\biggr)\widetilde{R}_{\s A}(\rho,p_{\rho})=0
\label{diciotto}
\end{equation}
%%%
So eq. (\ref{diciotto}) is the same as the free one 
(\ref{besselfree}) and as a consequence the spectrum $\widetilde{E}_{\s A}$ is the same as the 
$\widetilde{E}$ of (\ref{besselfree}). 
This is the proof that the spectrum of the Liouville operator {\it is not changed} by the presence 
of the gauge potential. Of course the two eigenfunctions which have the same eigenvalues $\widetilde{E}=
\widetilde{E}_{\s A}$ are different because they are labelled by different eigenvalues of the operator $\widehat{p}_{\varphi}$.
In fact the eigenfunction $\psi$ of eq. (\ref{psir}) has eigenvalue $p_{\varphi}^0$ for the operator
$\widehat{p}_{\varphi}$ while the eigenfunction $\psi_{\s A}$ of eq. (\ref{psir2}) has eigenvalue 
$\displaystyle p_{\varphi}^0+\frac{e\Phi_{\s B}}{2\pi c}$. 
So the two eigenfunctions are related by a shift in one of their "classical"
numbers $p^0_{\varphi}$.
The difference with the quantum case is that the corresponding eqs. (\ref{besselr}) and (\ref{besselmag}) cannot
be turned one into the other like we did in the KvN case. This is so because in (\ref{besselr})
$m$ is an integer and not a continuous real eigenvalue like $p^0_{\varphi}$ is in the KvN case.
Also in classical mechanics we had an integer eigenvalue, $n$ for $\displaystyle 
-i\frac{\partial}{\partial\varphi}$, but we managed
to down-load on $p^0_{\varphi}$, and not on $n$, the difference between the free and the interacting case.

The reader may object that, even if the classical spectrum is the same in the two cases, 
the eigenfunctions  are different and then
the evolution may lead to different results. Actually it is not so because, as we see from (\ref{sevensix}), we have to integrate
over  all the possible eigenvalues  
which label the eigenfunctions. In our case the different eigenfunctions (\ref{psir})-(\ref{psir2}) 
have only the "classical" number $p_{\varphi}^0$ shifted. Since $p_{\varphi}^0$
can assume every real number, when in
(\ref{sevensix}) we integrate over all the $p_{\varphi}^0$ a shift in them has no effect on the final result.

We feel that this proof that the spectrum of the classical KvN operator is unchanged by the presence of the gauge potential,
while the spectrum of the Schr\"odinger operator is changed, is the most convincing proof of the AB phenomenon. 
\section{Conclusions}

In this paper we have studied which is the minimal coupling procedure for the KvN operatorial approach to CM. We have 
shown that the MC involves not only the momenta but also their derivatives. We managed to encapsulate these two
MC into a single one using the concept of superfield. We have then applied this technique to the Landau problem
and to the Aharonov-Bohm phenomenon. In the first case (Landau problem) we showed that in the KvN 
formalism there
is a sort of discretization phenomenon in the eigenvalues of the evolution operator. Moreover we proved that
there are many more degeneracies in the 
classical than in the quantum case. For the second problem (the Aharonov-Bohm
one) we showed that at the quantum level there
is a change in the spectrum of the Schr\"odinger operator once the gauge potential is present while there is no change in 
the spectrum of the classical KvN operator. We feel this is the most convincing proof of the AB
effect. The paper contains also a complete analysis of the issue
of gauge invariance in the Hilbert space of KvN.

Having now all the tools to write down, in the KvN formalism, the interaction between a particle and a gauge field, what we
should do next is to see how the gauge fields interact among themselves 
in the KvN operatorial approach. This
has already been started and work is in progress on it \cite{Carta}.

\newpage
\begin{center}
{\LARGE\bf Appendices}
\end{center}

\appendix
\makeatletter
\@addtoreset{equation}{section}
\makeatother
\renewcommand{\theequation}{\thesection.\arabic{equation}}

\section{Appendix }

In this appendix we will briefly review the extended canonical formalism associated to the Hamiltonian $\HT$ of eq.
(\ref{supham}). From the Lagrangian (\ref{suplag}) one could derive the equations of motion for the $8n$-variables
$(\varphi^a,\lambda_a,c^a,\bar{c}_a)$ by the simple variational principle. These equations are:
%%%
\begin{eqnarray}
\label{eq:otto}
{\dot\varphi }^{a}-\omega^{ab}\partial_{b}H & = & 0 \\
\label{eq:nove}
[\delta^{a}_{b}\partial_{t}-\omega^{ac}\partial_{c}\partial_{b}H]c^{b}
 & = & 0 \\
\label{eq:dieci}
\delta^{a}_{b}\partial_{t}{\bar c}_{a}+{\bar
c}_{a}\omega^{ac}\partial_{c}\partial_{b}H & = & 0 \\
\label{eq:undici}
[\delta_{b}^{a}\partial_{t}+\omega^{ac}\partial_{c}\partial_{b}H]\lambda_{a}
& = & -i{\bar c}_{a}\omega^{ac}\partial_{c}\partial_{d}\partial_{b}H c^{d} 
\end{eqnarray}
%%%
We could ask ourselves if these same equations could be derived from the Hamiltonian $\HT$. The answer is yes. If we
introduce the following extended Poisson brackets structure in the extended space $(\varphi^a,\lambda_a,c^a,\bar{c}_a)$
%%%
\begin{eqnarray}
&&\{\varphi^a,\lambda_b\}_{epb}=\delta_b^a\\
&&\{\bar{c}_b,c^a\}_{epb}=-i\delta_b^a \label{epb}
\end{eqnarray}
%%%
(while all the other brackets are zero) we get that the equations of motion (\ref{eq:otto})-(\ref{eq:undici}) 
can be derived as
%%%
\begin{equation}
\frac{dO}{dt}=\{O,\HT\}_{epb}
\end{equation}
%%%
where $O$ is any of the variables $(\varphi^a,\lambda_a,c^a,\bar{c}_a)$ or any function of them.
More details can be found in ref.
\cite{Gozzi}.

\section{Appendix }

In this appendix we will prove that the Liouville eq. (\ref{lio2}) is invariant under the gauge transformations.
Let us write (\ref{lio2}) in the abstract form:
%%%
\begin{equation}
i\frac{d}{dt}|\psi(t)\rangle=\HCT|\psi(t)\rangle
\end{equation}
%%%
Let us now do a gauge transformation by a parameter $\alpha$. The new ket will be
%%% 
\begin{equation}
|\psi^{\prime}(t)\rangle=\widetilde{U}|\psi(t)\rangle
\end{equation}
%%%
where $\widetilde{U}$ is the expression in formula (\ref{uexp}).
We will  prove that this state satisfies the following equation
%%%
\begin{equation}
i\frac{d}{dt}|\psi^{\prime}(t)\rangle=\hat{\HT^{\prime}}|\psi^{\prime}(t)\rangle \label{btre}
\end{equation}
%%%
where $\hat{\HT^{\prime}}$ is the operator obtained from $\HCT$ by doing a gauge transformation
%%%
\begin{equation}
\hat{\HT^{\prime}}=\frac{1}{m}\biggl(\lambda_{q_i}+\frac{e}{c}\lambda_{p_k}
\partial_{q_k}A^{\prime}_{q_i}\biggr)\biggl(p_i-\frac{e}{c}A^{\prime}_{q_i}\biggr)-e\lambda_{p_i}\partial_{q_i}\Phi^{\prime}
\label{gaugeinvlio}
\end{equation}
%%%
where the $A_{q_i}^{\prime}$ is the gauge transformed vector potential and $\Phi^{\prime}$ is the gauge transformed scalar
potential. (\ref{gaugeinvlio}) is the $\HT$ in the gauge ($A^{\prime}$, $\Phi^{\prime}$) and it is derived from\break 
$\displaystyle H=\frac{(p_i-A_{q_i}^{\prime})^2}{2m}+e\Phi^{\prime}$. Let us now evaluate the LHS of (\ref{btre})
%%%
\begin{eqnarray}
\displaystyle
&&i\frac{d}{dt}|\psi^{\prime}(t)\rangle=i\frac{d}{dt}[\widetilde{U}(t)|\psi(t)\rangle]=i\biggl[\frac{d}{dt}\widetilde{U}(t)
\biggr]|\psi(t)\rangle+i\widetilde{U}(t)\frac{d}{dt}|\psi(t)\rangle=-\frac{e}{c}\frac{\partial\widetilde{\alpha}}{\partial
t}\widetilde{U}|\psi(t)\rangle\nonumber\\
&&+\widetilde{U}(t)\HCT|\psi(t)\rangle =-\frac{e}{c}\frac{\partial\widetilde{\alpha}}{\partial
t}|\psi^{\prime}(t)\rangle+\widetilde{U}(t)\HCT\widetilde{U}^{-1} |\psi^{\prime}(t)\rangle
=\biggl\{-\frac{e}{c}\frac{\partial\widetilde{\alpha}}{\partial t}+
\widehat{\cal H}\biggr\}|\psi^{\prime}(t)\rangle
\label{bcinque}
\end{eqnarray}
%%%
where 
%%%
\begin{equation}
\widehat{\cal H}\equiv\widetilde{U}\HCT\widetilde{U}^{-1} \label{bsei}
\end{equation}
%%%
In the first steps above we have used the expression (\ref{uexp}) for $\widetilde{U}$ and formula (\ref{aexp}). 
The explicit expression for $\widehat{\cal H}$ is
%%%
\begin{equation}
\widehat{\cal H}=\frac{1}{m}\biggl(\lambda_{q_i}^{\prime}+\frac{e}{c}\lambda_{p_k}^{\prime}\partial_{q_k}A_{q_j}
(q^{\prime})\biggr)\biggl(p_i^{\prime}-\frac{e}{c}A_{q_i}(q^{\prime})\biggr)-e\lambda_{p_i}^{\prime}\partial_{q_i}
\Phi(q^{\prime})
\label{tildetilde}
\end{equation}
%%%
where
%%%
\begin{equation}
\left\{
	\begin{array}{l}
	\displaystyle 
	\lambda_{q_i}^{\prime}=\widetilde{U}\lambda_{q_i}\widetilde{U}^{-1}=\lambda_{q_i}+\frac{e}{c}\lambda_{p_k}
          \partial_{q_k}\partial_{q_i}\alpha(q)\smallskip\\
          \displaystyle p_i^{\prime}=\widetilde{U}p_i\widetilde{U}^{-1}=p_i-\frac{e}{c}\partial_{q_i}\alpha(q)\smallskip\\
          \displaystyle q_i^{\prime}=\widetilde{U}q_i\widetilde{U}^{-1}=q_i\smallskip\\
          \displaystyle \lambda_{p_i}^{\prime}=\widetilde{U}\lambda_{p_i}\widetilde{U}^{-1}=\lambda_{p_i}\\
	\end{array}
	\right.
\end{equation}
%%% 
Remembering that 
%%%
\begin{equation}
\left\{
	\begin{array}{l}
	\displaystyle A^{\prime}_{q_i}=A_{q_i}+\partial_{q_i}\alpha\smallskip\\
          \displaystyle \Phi^{\prime}=\Phi-\frac{1}{c}\frac{\partial\alpha}{\partial t}\\
          \end{array}
	\right.
\end{equation}
%%%
we can rewrite the $\widehat{\cal H}$ of (\ref{tildetilde}) as
%%%
\begin{eqnarray}
\widehat{\cal H}&=&\frac{1}{m}\biggl(\lambda_{q_i}+\frac{e}{c}\lambda_{p_k}\partial_{q_k}\partial_{q_i}\alpha(q)
+\frac{e}{c}\lambda_{p_k}\partial_{q_k}A_{q_i}(q)\biggr)
\biggl(p_i-\frac{e}{c}\partial_{q_i}\alpha(q)-\frac{e}{c}A_{q_i}(q)\biggr)\nonumber\\
&&-e\lambda_{p_i}\partial_{q_i}\Phi
=\frac{1}{m}\biggl(\lambda_{q_i}+\frac{e}{c}\lambda_{p_k}\partial_{q_k}A^{\prime}\biggr)\biggl(p_i-\frac{e}{c}
A_{q_i}^{\prime}
\biggr)-e\lambda_{p_j}\partial_{\partial j}\biggl[\Phi^{\prime}(q)+\frac{1}{c}\frac{\partial
\alpha}{\partial t}\biggl]=\nonumber\\
&&=\hat{\HT^{\prime}}+\frac{e}{c}\frac{\partial}{\partial t}\widetilde{\alpha}
\end{eqnarray}
%%%
Using this result (\ref{bcinque}) becomes
%%%
\begin{equation}
i\frac{d}{dt}|\psi^{\prime}\rangle=\hat{\HT^{\prime}}|\psi^{\prime}\rangle
\end{equation}
%%%
which is what we wanted to prove.

\section{Appendix }

We know that one of the effects of the introduction of the MC is that local phases multiplying the states can be absorbed by
a gauge transformation of the gauge field. What we mean is the following: if 
%%%
\begin{equation}
\displaystyle \psi^{\prime}(q)=exp\biggl[i\frac{e}{c\hbar}\alpha(q)\biggr] \label{cuno}
\psi(q)
\end{equation}
%%% 
and
\begin{equation}
\displaystyle \widehat{H}=-\frac{1}{2m}\biggl(-i\hbar\frac{\partial}{\partial
q}-\frac{e}{c}A(q)\biggr)\biggl(-i\hbar\frac{\partial}{\partial q} 
-\frac{e}{c}A(q)\biggr)
\end{equation} then 
%%%
\begin{equation}
\widehat{H}^{\prime}\psi^{\prime}=exp\biggl[i\frac{e}{c\hbar}\alpha(q)\biggr]\widehat{H}\psi \label{cdue}
\end{equation}
%%%
where $\widehat{H}^{\prime}$ is obtained from  $\widehat{H}$ by replacing $\widehat{A}$ with its gauge transformed
$\displaystyle A^{\prime}=A+\frac{\partial\alpha}{\partial q}$. Now if we do the minimal coupling at the level of $\HCT$, like in
(\ref{abstham}), the phase that can "pass through", like in (\ref{cdue}), is $
\displaystyle exp\biggl[i\frac{e}{c}\widetilde{\alpha}\biggr]$ with
$\widetilde{\alpha}$ given in (\ref{aexp}):
%%%
\begin{equation}
\displaystyle
\hat{\HT^{\prime}}\,\psi^{\prime}=exp\biggl[i\frac{e}{c}\widetilde{\alpha}\biggr]\,\HCT\,\psi \label{ctre}
\end{equation}
%%%
The proof goes as follows. Let us use the mixed representation (\ref{mixrep}) for the $\HCT$ associated to
(\ref{abstham}):
%%%
\begin{equation}
\displaystyle
\HCT=\frac{1}{m}\biggl(-i\frac{\partial}{\partial q}+\frac{e}{c}\lambda_p\frac{\partial A}{\partial q}\biggr)
\biggl(i\frac{\partial}{\partial \lambda_p}-\frac{e}{c}A(q)\biggr) \label{cquattro}
\end{equation}
%%%
and let us then transform $\HCT$ into a $\hat{\HT^{\prime}}$ where
$\displaystyle A\rightarrow A^{\prime}+\frac{\partial\alpha}{\partial q}$.
In the same mixed representation (\ref{mixrep}) the wave function will be of the form $\psi(q,\lambda_p)$ 
and we can also construct  the following new state:
%%%
\begin{equation}
\displaystyle
\psi^{\prime}(q,\lambda_p)=exp\biggl[-i\frac{e}{c}\lambda_p\partial_q\alpha(q)\biggr]\psi(q,\lambda_p)=
exp\biggl[i\frac{e}{c}\widetilde{\alpha}\biggr]\psi(q,\lambda_p)
\label{ccinque}
\end{equation}
%%%
Equipped with these tools it is then a long but easy calculation to prove (\ref{ctre}). To conclude we
can say that the local phase
transformations of the form (\ref{ccinque}) on the KvN states $\psi(q,\lambda_p)$ are the {\it classical} counterpart of
the  local phase transformations (\ref{cuno}) on the quantum Hilbert states $\psi(q)$. 

At this point a question which arises naturally is  the 
following: if instead of the very particular phase transformation (\ref{ccinque}) on the KvN states we perform a general local
phase transformation of the form
%%%
\begin{equation}
\psi^{\prime}(q,\lambda_p)=exp[i\alpha(q,\lambda_p)]\psi(q,\lambda_p)
\end{equation}
%%%
which gauge fields do we have to introduce in the $\HCT$ in order to absorb the phase like in (\ref{ctre})?
The answer is the following: let us start from the $\HCT$ of the free particle
%%%
\begin{equation}
\HCT=\frac{1}{m}\frac{\partial^2}{\partial q\partial\lambda_p} \label{csei}
\end{equation}
%%%
and do a general phase transformation 
%%%
\begin{equation}
\displaystyle
\psi^{\prime}(q,\lambda_p)=exp[i\alpha(q,\lambda_p)]\psi(q,\lambda_p) \label{cseib}
\end{equation}
%%%
If we now perform  in (\ref{csei}) the following MC 
%%%
\begin{eqnarray}
\displaystyle
&&-i\frac{\partial}{\partial q}\;\longrightarrow\;-i\frac{\partial}{\partial
q}+A_q,\;\;\;\;\;\;\;\;\;i\frac{\partial}{\partial\lambda_p}\;\longrightarrow\;i\frac{\partial}{\partial\lambda_p}-A_{\lambda_p}
\end{eqnarray}
%%%
where $A_q$ and $A_{\lambda_p}$ are two gauge fields which transform as follows
%%%
\begin{eqnarray}
&&A_q^{\prime}=A_q-\frac{\partial\alpha(q,\lambda_p)}{\partial
q},\;\;\;\;\;\;A_{\lambda_p}^{\prime}=A_{\lambda_p}-\frac{\partial\alpha(q,\lambda_p)}{\partial\lambda_p} \label{cotto}
\end{eqnarray}
%%%
then the new $\HCT_{\s A}$:
%%%
\begin{equation}
\displaystyle
\HCT_{\s A}\equiv\frac{1}{m}\biggl(-i\frac{\partial}{\partial
q}+A_q\biggr)\biggl(i\frac{\partial}{\partial\lambda_p} -A_{\lambda_p}\biggr) \label{cnove}
\end{equation}
%%%
would satisfy the following relation
%%%
\begin{equation}
\HCT^{\prime}_{\s A}\psi^{\prime}=exp[i\alpha(q,\lambda_p)]\HCT_{\s A}\psi
\end{equation}
%%%
where $\HT^{\prime}_{\s A}$ is the gauge transformed of $\HCT_{\s A}$ via eq. (\ref{cotto}). 

We should notice that (\ref{cseib})
is a more general gauge transformation than the one in (\ref{ccinque}). To implement  (\ref{cseib}) we need two gauge fields $A_q$,
$A_{\lambda_p}$ while for the (\ref{ccinque}) we could build everything from one field $A(q)$, see (\ref{cquattro}). The 
transformation (\ref{ccinque}) is a particular case of (\ref{cseib}) and the same is true for the gauge fields which enter
(\ref{cquattro}). In fact comparing (\ref{cquattro}) with (\ref{cnove}) we see that the two general gauge fields of (\ref{cnove}) 
should  be of the following particular form in order to reproduce (\ref{cquattro}):
%%%
\begin{equation}
A_q=\frac{e}{c}\lambda_p\frac{\partial A}{\partial q},
\;\;\;\;\;\;\;A_{\lambda_p}=\frac{e}{c}A(q) \label{cdieci}
\end{equation}
%%%
It is interesting to notice that the interaction introduced by the two general gauge fields $A_q,A_{\lambda_p}$ appearing in
(\ref{cnove}) is something new and not related to any magnetic field in $q$-space unless the fields 
have the particular form (\ref{cdieci}). We hope to come back in the future to the study of the forces generated by the general
gauge potential of (\ref{cnove}).

\section{Appendix }

In this appendix we want to build and diagonalize the $\HCT$ associated to an harmonic oscillator in one dimension. The
Hamiltonian is
%%%
\begin{equation}
H=\frac{1}{2m}p^2+\frac{1}{2}m\omega^2q^2
\end{equation}
%%%
So the $\HT$  (\ref{hambos}) is :
%%%
\begin{equation}
\HT=\lambda_q\frac{p}{m}-m\omega^2\lambda_pq
\end{equation}
%%%
which, in its operatorial version, using the representation (\ref{oprep}), is:
%%%
\begin{equation}
\displaystyle
\HCT=-i\biggl(\frac{p}{m}\frac{\partial}{\partial q}-m\omega^2q\frac{\partial}{\partial p}\biggr) \label{ditre}
\end{equation}
%%%
This operator is very similar to the components of the angular momentum in standard quantum mechanics, so we can 
diagonalize it with similar techniques. First of all we can turn the standard phase space coordinates $(q,p)$ into
the following new ones $(r,\theta)$ defined as:
%%%
\begin{equation}
\displaystyle
\sqrt{m}\omega q=r cos\theta,\;\;\;\;\;\;\frac{p}{\sqrt{m}}=r sin\theta
\end{equation}
%%%
The Hamiltonian becomes:
%%%
\begin{equation}
\displaystyle 
\HCT=-i\frac{r}{\sqrt{m}}sin \theta\biggl(\frac{\partial r}{\partial q}\frac{\partial}{\partial r}+\frac{\partial \theta}{\partial q}
\frac{\partial}{\partial \theta}\biggr)+i\sqrt{m}\omega r cos \theta\biggl(\frac{\partial r}{\partial p}
\frac{\partial}{\partial r}+\frac{\partial\theta}{\partial p}\frac{\partial}{\partial\theta}\biggr)
=i\omega\frac{\partial}{\partial \theta}
\end{equation}
The eigenfunctions of the previous operator are:
\begin{equation}
\displaystyle
\psi(r,\theta)=F(r)e^{-iN\theta}
\end{equation}
and the associated eigenvalues are $\widetilde{E}=N\omega$. Imposing the single valuedness of the wave functions 
we get that
%%%
\begin{equation}
\displaystyle 
\psi(r,\theta+2\pi)=\psi(r,\theta) \;\Rightarrow \;e^{-2i\pi N}=1 \label{disette}
\end{equation}
%%%
which implies: $N\in\{0,\pm1,\pm2,\dots\}$. So the discretization of the eigenvalues of the Liouvillian is a direct 
consequence of the requirement of single valuedness of the KvN wave functions, requirement that was already present in 
Koopman's original paper \cite{Koopman}.

The proof of the discretization of the spectrum of the Liouville operator for an harmonic oscillator can be worked out 
also in the mixed representation (\ref{mixrep}) and (\ref{diagmix}). In this representation the $\HT$ of eq.
(\ref{hambos}) becomes:
%%%
\begin{equation}
\HCT=\frac{1}{m}\frac{\partial}{\partial q}\frac{\partial}{\partial\lambda_p}-m\omega^2\lambda_pq \label{dtre}
\end{equation}
%%%
Let us introduce  the following new variables
%%%
\begin{equation}
Z_+\equiv\frac{q+\Delta\lambda_p}{\sqrt{2}},\;\;\;\;\;\;\;\;Z_-\equiv\frac{q-\Delta\lambda_p}{\sqrt{2}} \label{paolo}
\end{equation}
%%%
where $\Delta$ is a constant which has the dimension of an action. In terms of these new variables the $\HCT$ of
(\ref{dtre}) can be written as
%%%
\begin{eqnarray}
\displaystyle
\HCT&=&\frac{1}{\Delta}\biggl[-\frac{\Delta^2}{2m}\frac{\partial^2}{\partial
Z_-^2}+\frac{m\omega^2}{2}Z_-^2\biggr] -\frac{1}{\Delta}\biggl[-\frac{\Delta^2}{2m}\frac{\partial^2}{\partial
Z_+^2}+\frac{m\omega^2}{2}Z_+^2\biggr]=\nonumber\\ &=&\frac{1}{\Delta}\biggl[H^{osc}\biggl(Z_-,\frac{\partial}{\partial
Z_-}\biggr)-H^{osc}\biggl(Z_+,\frac{\partial}{\partial Z_+}\biggr)\biggr]
\end{eqnarray}
%%%
As indicated in the second step above, we notice that $\HCT$ is the difference of two quantum harmonic oscillators
respectively in $Z_-$ and $Z_+$, where the role of $\hbar$ is taken by our constant $\Delta$.
The eigenstates of $\HCT$ :
%%%
\begin{equation}
\HCT\psi(Z_+,Z_-)=\widetilde{E}\,\psi(Z_+,Z_-)
\end{equation}
%%%
can  be easily obtained. They are $\psi(Z_+,Z_-)=\psi_n^{osc}(Z_+)\psi_m^{osc}(Z_-)$ with 
%%%
\begin{equation}
\displaystyle
\psi_n(Z_{\pm})=(\sqrt{\pi}2^nn!\sigma_0)^{-1/2}H_n\biggl(\frac{Z_{\pm}}{\sigma_0}\biggr)
exp\biggl(-\frac{Z_{\pm}^2}{2\sigma_0^2}\biggr),\;\;\;\;n=0,+1,+2,\cdots
\end{equation}
%%%
where $H_n$ are the Hermite polynomials and $\sigma_0=\sqrt{\frac{\Delta}{m\omega}}$.
The eigenvalues are:
%%%
\begin{equation}
\widetilde{E}_{n,m}=\frac{1}{\Delta}\biggl[\biggl(m+\frac{1}{2}\biggr)\Delta\omega-\biggl(n+\frac{1}{2}\biggr)
\Delta\omega\biggl]=(m-n)\omega=N\omega \label{dsei}
\end{equation}
%%%
where $N$ can take every positive or negative integer value: $N=0,\pm 1,\pm 2,\cdots$.
This confirms the discretization phenomenon we found before. Let us notice that the quantity $\Delta$ disappears 
from the spectrum, so it is just an artifact of the $(q,\lambda_p)$ representation and it is needed in (\ref{paolo})
only for dimensional reasons. There was no need of it in the first derivation, eqs. (\ref{ditre})-(\ref{disette}),
of the discretization phenomenon.
 Second let us notice that, due to the difference of the two oscillators quantum numbers $m$ and $n$ above, 
the zero-point "energy"\footnote{We have put quotation marks around the word "energy"
because, as we explained in section 4, the $\widetilde{E}$ is not
the energy but one of the eigenvalues of the evolution operator.} is zero 
differently than in the quantum case. 
Note also that there is
an $\infty$-order degeneracy in the sense that associated to the eigenstate $\widetilde{E}=N\omega$ there is the 
set of eigenfunctions:
$\psi=\psi_n(Z_+)\psi_{n+N}(Z_-)$, where $n$ can be any integer if $N\ge 0$ while $n>-N$ if $N<0$.

This doubling of oscillators in the classical case is basically due to the fact that the classical KvN wave functions 
depend on a
number of variables $(q,\lambda_p)$ which is double with respect to the quantum case. 

\section{Appendix }

In section 4 we have analyzed the properties of the Landau problem in the Schr\"odinger picture
of the KvN formalism. In this appendix we want to study the same problem
in the corresponding Heisenberg picture\footnote{All the objects appearing in this appendix are
operators. Therefore we will not use explicitly the hat-symbol $("\wedge")$ to indicate them.}. 
In particular we want to find out which are the constants of motion, i.e. 
the operators that commute with the generator of the time evolution ${\mathcal H}$. 
These operators will give us some indications
concerning the trajectory of the classical particle in a constant magnetic field. 

Let us remember the form of the Liouvillian in the Landau problem:
%%%
\begin{equation}
\displaystyle
{\cal H}=\frac{1}{m}\lambda_xp_x+\frac{1}{m}\biggl(\lambda_y+\frac{e}{c}B\lambda_{p_x}\biggr)
\biggl(p_y-\frac{eB}{c}x\biggr)+\frac{1}{m}\lambda_zp_z
\end{equation}
%%%
Defining $\displaystyle v_y=\frac{1}{m}\biggl(p_y-\frac{e}{c}A_y\biggr)$ and noticing
that $[x,{\cal H}]=ip_x/m$,
we get the following relation:
%%%
\begin{equation}
\displaystyle
[v_y,{\cal H}]=\frac{1}{m}\biggl[p_y-\frac{e}{c}A_y,{\cal H}\biggr]=-\frac{e}{m^2c}[Bx,\lambda_x]p_x
=-\frac{ieB}{m^2c}p_x \label{cos1}
\end{equation}
%%%
If we introduce the Larmor frequency: $\displaystyle \omega=\frac{eB}{mc}$ we can then easily prove
that\break $\displaystyle x_0\equiv x+v_y/\omega$ is a constant of motion.
In fact, using (\ref{cos1}) we get:
%%%
\begin{equation}
[x_0,{\cal H}]=\biggl[x+\frac{v_y}{\omega},{\cal H}\biggr]=\frac{i}{m}p_x+\frac{mc}{eB}\cdot
\biggl(-\frac{ieB}{m^2c}p_x\biggr)=0
\end{equation}
%%%
In the same way the commutators of $y$ and $v_x$ with the Liouvillian are:
%%%
\begin{equation}
\displaystyle
[y,{\cal H}]=\frac{i}{m}\biggl(p_y-\frac{eB}{c}x\biggr),\;\;\;\;\;\;\;
[v_x,{\cal H}]=\frac{ieB}{m^2c}\biggl(p_y-\frac{eB}{c}x\biggr) \label{cos2}
\end{equation}
%%%
and so we obtain that also $\displaystyle y_0\equiv y-v_x/\omega$ commutes with 
${\mathcal H}$:
%%%
\begin{equation}
[y_0,{\cal H}]=[y,{\cal H}]-\frac{mc}{eB}[v_x,{\cal H}]=0
\end{equation}
%%%
Now classically a particle in a constant magnetic field
directed along $z$ describes an helicoidal orbit whose projection 
on the $x,y$-plane is a circumference with a radius equal to the Larmor one $\varrho_{Lar}$:
%%%
\begin{equation}
\displaystyle
\varrho_{Lar}^2\equiv \frac{1}{\omega^2}(v_x^2+v_y^2)
\end{equation}
%%%
Using eqs. (\ref{cos1})-(\ref{cos2}) it is possible to prove that also the 
Larmor radius is a constant of the motion:
%%%
\begin{equation}
\displaystyle
[\varrho_{Lar}^2,{\cal H}]=\frac{1}{\omega^2}[v_x^2,{\cal H}]+\frac{1}{\omega^2}[v_y^2,{\cal H}]=0
\end{equation}
%%%
The Larmor radius can be written also in terms of the $x, x_0$ and $y,y_0$ operators in the following way:
%%%
\begin{equation}
\displaystyle
\varrho^2_{Lar}=\frac{1}{\omega^2}(v_x^2+v_y^2)=\frac{1}{\omega^2}[\omega(y-y_0)]^2+
\frac{1}{\omega^2}[\omega(x_0-x)]^2=(x-x_0)^2+(y-y_0)^2
\end{equation}
%%%
Therefore $(x_0,y_0)$ is the center of a circumference which is the projection of the orbit of the particle
onto the plane $(x,y)$ and $\varrho_{Lar}$ is the corresponding radius.
Note that in the KvN operatorial formalism the operators $x_0$ and $y_0$ are suitable combinations
of the $\varphi$ operators and they commute among themselves. This implies that they can be determined 
with arbitrary precision. In quantum mechanics, instead, one can prove that the following relation holds:
%%%
\begin{equation}
[x_0,y_0]=\frac{-i\hbar c}{eB}
\end{equation}
%%%
and therefore, differently than in classical mechanics, 
there is an uncertainty relation involving the coordinates of 
the center of the circumference.

\section*{Acknowledgments}

We wish to thank F. Benatti for asking, long ago, some questions which triggered the present investigation. This work has
been supported by grants from INFN, MURST and the University of Trieste.

\newpage
\noindent {\LARGE \underline{Figure Caption}}

\bigskip
\bigskip
\bigskip

\noindent {\large Figure  1}: Aharonov-Bohm geometrical set up.\\

\noindent {\large Figure 2}: Zeros of Bessel functions: m=1 (continuous line),
m=0.9 (dashed line).

\newpage

\begin{thebibliography}{99}
\bibitem{Koopman}
B.O. Koopman, Proc. Natl. Acad. Sci. U.S.A. {\bf 17}, 315 (1931);\\
J. von Neumann, {\it Ann. Math.} {\bf 33}, 587 (1932); ibid. {\bf 33}, 789 (1932);
\bibitem{Mauro}
D. Mauro, {\it "On Koopman-von Neumann Waves"}, quant-ph/0105112;
\bibitem{quantumbooks}
J.J. Sakurai, {\it Modern Quantum Mechanics}, Rev.Ed.,
Addison-Wesley, Reading (MA) 1995;\\
C. Cohen Tannoudji et al., {\it Quantum Mechanics}, Wiley, New York , 1977;
\bibitem{Gozzi}
E. Gozzi, M. Reuter and W.D. Thacker, Phys. Rev. D {\bf 40} 3363 (1989);\\
E. Deotto, G. Furlan and E. Gozzi, Jour. Math. Phys. {\bf 41}, 8083 (2000);
\bibitem{Regini}
E. Gozzi, M. Regini, Phys. Rev. D {\bf 62}, 067702 (2000);\\
E. Gozzi, D. Mauro, Jour. Math. Phys. {\bf 41}, 1916 (2000);
\bibitem{metaplectic}
E. Gozzi, M. Reuter, Jour. Phys. A {\bf 26}, 6319 (1995);\\
E. Deotto, E. Gozzi and D. Mauro, work in progress;
\bibitem{Abrikosov}
A.A. Abrikosov and E. Gozzi, Nucl. Phys. B Proc. Suppl. {\bf 88}, 369 (2000) 
(quant-ph/9912050);\\
A.A. Abrikosov, E. Gozzi and D. Mauro, work in progress;
\bibitem{Bohm}
Y. Aharonov, D. Bohm, Phys. Rev. {\bf 115}, 485 (1959);
\bibitem{Watson}
E.T. Whittaker, G.N. Watson, {\it A Course of Modern Analysis}, Cambridge University Press (1973);
\bibitem{Carta}
P. Carta, Master Thesis, Cagliari University, 1994;\\
D. Mauro, Master Thesis, Trieste University, 1999;\\
P. Carta, E. Gozzi and D. Mauro, work in progress.


\end{thebibliography}
\end{document}